\documentclass[aps,pra,reprint,twocolum,amsmath,superscriptaddress,amssymb,showpacs]{revtex4-1}
\usepackage{amsmath,braket,amssymb}
\usepackage[final]{graphicx}
\usepackage{subfigure}
\usepackage{xcolor}
\usepackage[english]{babel}
\usepackage[utf8]{inputenc}
\usepackage{color}
\usepackage{mathtools}
\usepackage{empheq}
\usepackage{tensor}

\usepackage[colorlinks,citecolor=darkBlue,linkcolor=darkBlue,
urlcolor=blue,hyperindex]{hyperref}

\usepackage[normalem]{ulem}
\normalfont

\definecolor{forestgreen}{rgb}{0.08, 0.4, 0.13}
\definecolor{darkBlue}{rgb}{0.08, 0.13, 0.4}

\definecolor{THc}{rgb}{0.9,0.3,0.2}

\usepackage{amsthm}
\theoremstyle{definition}

\theoremstyle{plain}

\newcommand{\prlsection}[1]{{\em {#1}.---~}}

\begin{document}
	
\title{Quantifying Nonstabilizerness of Matrix Product States}
	
\author{Tobias Haug}
\email{tobias.haug@u.nus.edu}
\affiliation{QOLS, Blackett Laboratory, Imperial College London SW7 2AZ, UK}

\author{Lorenzo Piroli}
\affiliation{Philippe Meyer Institute, Physics Department, \'{E}cole Normale Sup\'{e}rieure (ENS), Universit\'{e} PSL, 24 rue Lhomond, F-75231 Paris, France}

\date{\today}
	
\begin{abstract}
Nonstabilizerness, also known as magic, quantifies the number of non-Clifford operations needed in order to prepare a quantum state. As typical measures either involve minimization procedures or a computational cost exponential in the number of qubits $N$, it is notoriously hard to characterize for many-body states. In this work, we show that nonstabilizerness, as quantified by the recently introduced Stabilizer R\'enyi Entropies (SREs), can be computed efficiently for matrix product states (MPSs). Specifically, given an MPS of bond dimension $\chi$ and integer R\'enyi index $n>1$, we show that the SRE can be expressed in terms of the norm of an MPS with bond dimension $\chi^{2n}$. For translation-invariant states, this allows us to extract it from a single tensor, the transfer matrix, while for generic MPSs this construction yields a computational cost linear in $N$ and polynomial in $\chi$. We exploit this observation to revisit the study of ground-state nonstabilizerness in the quantum Ising chain, providing accurate numerical results up to large system sizes.  We analyze the SRE near criticality and investigate its dependence on the local computational basis, showing that it is in general not maximal at the critical point.
\end{abstract}
	
\maketitle


\section{Introduction} 
The very idea of quantum computers owes its origin to the difficulty of simulating quantum many-body physics on a classical one~\cite{feynman1982simulating}. Yet, there exist classes of quantum states which can be simulated classically. A prominent example is that of \emph{stabilizer states}, \emph{i.e.} the states generated by Clifford operations~\cite{gottesman1997stabilizer,gottesman1998theory,gottesman1998heisenberg,aaronson2004improved}.

Cliffords are an important class of unitaries in quantum information theory~\cite{nielsen2011quantum}. They can be implemented fault-tolerantly~\cite{shor1996fault,preskill1998fault} in many prototypical error-correcting quantum codes~\cite{kitaev2003fault,eastin2009restriction}, allowing for universal computation if supplemented with suitable nonstabilizer ancillary states~\cite{bravyi2005universal,campbell2017roads}. They also play a prominent role in many-body physics, as building blocks to construct tractable toy models for, \emph{e.g.}, non-equilibrium entanglement dynamics~\cite{nahum2017quantum} or the AdS/CFT correspondence~\cite{pastawski2015holographic}. 

An important task is to quantify the degree to which a quantum state can not be prepared by Clifford gates. This property, called \emph{nonstabilizerness} or magic~\cite{kitaev2003fault}, is related to the difficulty of classically simulating quantum states~\cite{howard2014contextuality,bravyi2016trading,bravyi2019simulation,seddon2021quantifying,koukoulekidis2022born}, and has been argued to be a necessary condition for quantum chaos~\cite{leone2022stabilizer,leone2021quantum,haferkamp2022random}. 

It was recently suggested that nonstabilizerness is an interesting quantity in many-body settings, shedding light, for instance, on the structure of ground-state (GS) wave-functions~\cite{white2021conformal,sarkar2020characterization,sewell2022mana,oliviero2022magic,liu2022many}. In particular, an intriguing connection was put forward between criticality and ``long-range magic''~\cite{white2021conformal,sarkar2020characterization,sewell2022mana}, \emph{i.e.} magic which can not be removed by quantum circuits of finite depth~\cite{sewell2022mana}. Unfortunately, measures of nonstabilizerness are typically hard to compute~\cite{campbell2011catalysis,veitch2014resource,howard2017application,wang2019quantifying,beverland2020lower,jiang2021lower,hahn2022quantifying, liu2022many,bu2022complexity}, especially when the local Hilbert space dimension is even~\cite{campbell2012magic,anwar2014fast,campbell2014enhanced}. While an efficient measurement protocol for quantum computers has been recently demonstrated~\cite{haug2022scalable}, quantitative investigations of these ideas remain difficult.

In this context, useful measures of magic, the Stabilizer R\'enyi Entropies (SREs), were recently introduced in Ref.~\cite{leone2022stabilizer}. They are expressed in terms of the expectation values of all Pauli strings and allow for explicit computations as exemplified in Ref.~\cite{oliviero2022magic} for the GS of the transverse-field Ising model. They can be probed experimentally by randomized measurement protocols~\cite{oliviero2022measuring} or Bell measurements~\cite{haug2022scalable}. However, the computational cost to evaluate the SRE of generic states grows exponentially in the number of qubits $N$, strongly limiting the system sizes which can be studied.

\begin{figure}
	\includegraphics[scale=0.16]{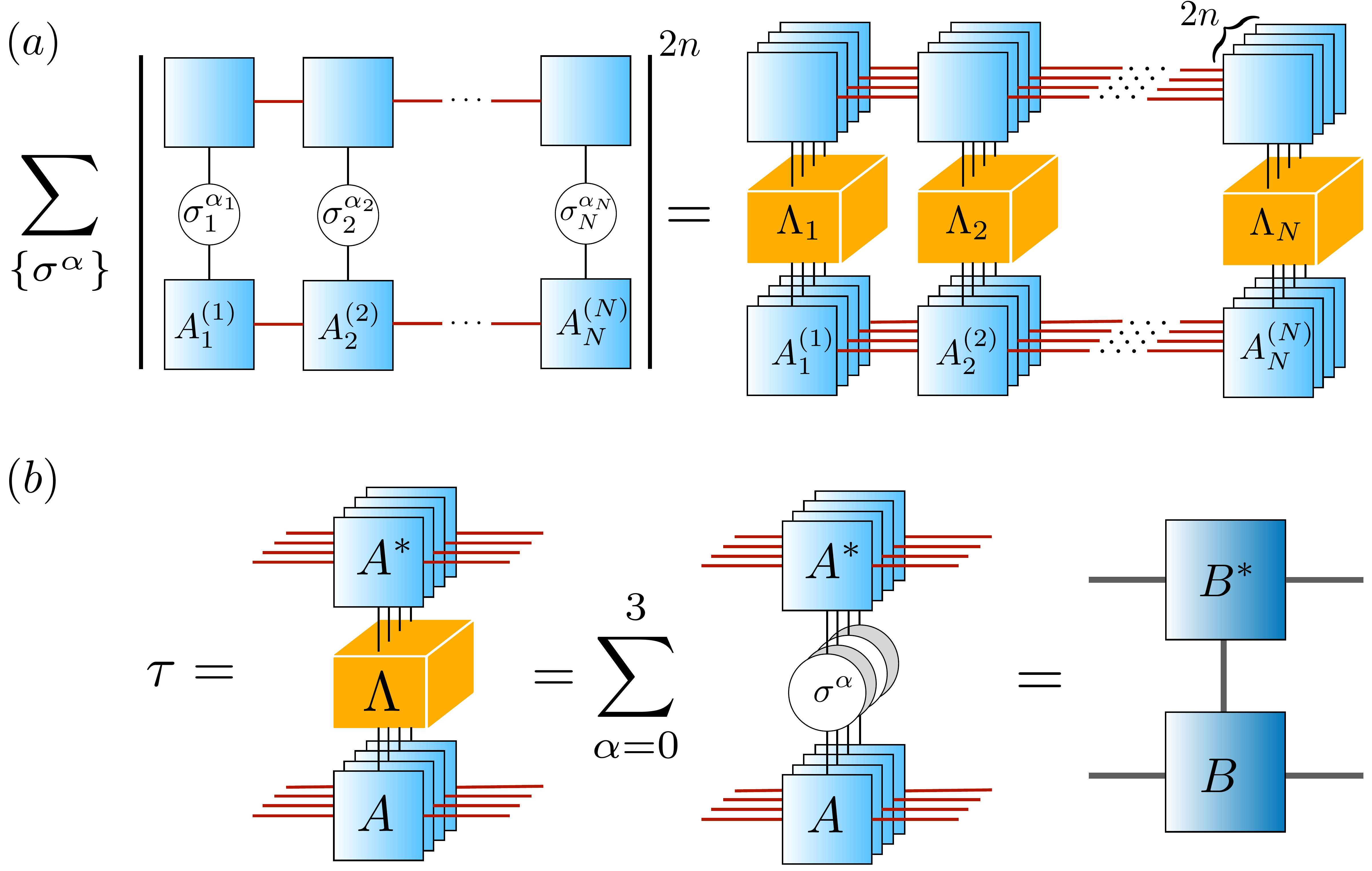}
	\caption{$(a)$ Pictorial representation of the replica approach, and the contraction identity~\eqref{eq:contraction}, for integer R\'enyi index $n$. $(b)$ The transfer matrix $\tau$, encoding full information on the stabilizer R\'enyi-$n$ entropy for TI MPSs.}
	\label{fig:sketch}
\end{figure}

Here, we show that, for integer R\'enyi index $n$, the SREs can be computed efficiently for the important class of Matrix Product States (MPSs)~\cite{perez2007matrix,cirac2017matrix_op,cirac2020matrix}. By mapping the SRE onto the norm of a certain MPS, cf. Fig.~\ref{fig:sketch}, we show that magic can be extracted out of a single tensor for translation-invariant (TI) states, while it can be computed at a cost linear in $N$ for generic MPSs. Based on this result, we revisit the study of magic in the quantum Ising model for large $N$. We analyze the SRE near criticality and investigate its dependence on the local computational basis. 

The rest of this work is organized as follows. In Sec.~\ref{sec:SRE_MPS} we explain the main idea to compute the SRE in MPSs. We show how it can be computed locally for TI MPSs, and discuss the efficient numerical procedure for its evaluation in the general, non-TI case. These results are applied in Sec.~\ref{sec:SRE_QIM}, which contains our study of the quantum Ising chain, while our conclusions are consigned to Sec.~\ref{sec:outlook}. Finally, the most technical part of our work, together with additional numerical results, are reported in the Appendix.
 
\section{SRE and Matrix Product States} 
\label{sec:SRE_MPS}
\subsection{Preliminaries}
We consider a system of $N$ qubits, with Hilbert space $\mathcal{H}=\otimes_{j=1}^N\mathcal{H}_j$, and $\mathcal{H}_j\simeq \mathbb{C}^{2}$. We denote by $\{\sigma^\alpha\}_{\alpha=0}^3$ the Pauli matrices ($\sigma^{0}=\openone$), by $\mathcal{P}_N$ the set of all $N$-qubit Pauli strings, and by $\{\ket{0}$, $\ket{1}\}$ the local computational basis. Given a pure (normalized) state $\ket{\Psi_N}\in\mathcal{H}$, the SRE of order $n$ reads~\cite{leone2022stabilizer}
\begin{equation}\label{eq:SRE}
M^{(n)}(|\Psi_N\rangle)=(1-n)^{-1} \ln \sum_{P \in \mathcal{P}_{N}} \frac{\braket{\Psi_N|P|\Psi_N}^{2n}}{2^N}\,.
\end{equation}
The SRE is a measure of nonstabilizerness in the following sense~\cite{leone2022stabilizer}: (i) it is zero iff $\ket{\Psi_N}$ is a stabilizer state; (ii) it is invariant under Clifford unitaries; (iii) it is additive under tensor product. We will consider the case where $\ket{\Psi_N}$ is an MPS~\cite{perez2007matrix,cirac2017matrix_op,cirac2020matrix}
\begin{equation}\label{eq:MPS}
\ket{\Psi_{N}}=\sum_{\{s_{k}\}} {\rm tr}\left(S A^{s_{1}}_1 \ldots A^{s_{N}}_N\right)\ket{s_{1}, \ldots, s_{N}}\,,
\end{equation}
where $A^{s}_{k}$ are $\chi\times \chi$ matrices. We call $\chi$ the \emph{bond dimension}, as opposed to the physical local dimension $d$ ($d=2$ for qubits). If $S=\ket{R}\bra{L}$, $\ket{\Psi_{N}}$ is an MPS with open boundary conditions (OBCs), while if $S=\openone$, and $A^{s}_j=A^{s}_k$ $\forall j,k$ we say that $\ket{\Psi_{N}}$ is a TI MPS with periodic boundary conditions (PBCs). In this case, we will further assume that $A_k$ are \emph{normal}~\cite{cirac2017matrix_op}. This is a technical condition, ensuring that $\ket{\Psi_N}$ does not have long-range correlations. Note that the state~\eqref{eq:MPS} is not necessarily normalized.

MPSs admit a useful graphical representation~\cite{cirac2017matrix_op}, where each matrix $A^{s}_k$ is interpreted as a tensor with three indices, denoted by three outer legs, cf. Fig.~\ref{fig:sketch}. Legs shared by two tensors, $A$ and $B$, correspond to a \emph{contraction}, meaning that the associated common index is summed over~\cite{orus2014practical}. We will denote by $A\cdot B$ the tensor obtained by contracting the legs shared by $A$ and $B$. 

MPSs are an invaluable tool in one-dimensional many-body physics, representing faithfully GSs of local Hamiltonians~\cite{verstraete2006matrix,schuch2008entropy} and being at the basis of powerful numerical algorithms~\cite{schollwock2011density}. For any MPS $\ket{\Psi_N}$ and $P\in\mathcal{P}_N$, the expectation values $\braket{\Psi_N|P|\Psi_N}$ can be computed efficiently, \emph{i.e.} at a cost linear in $N$. Yet, since the SRE involves a sum of $4^N$ terms, a straightforward evaluation of Eq~\eqref{eq:SRE} results in a cost exponential in $N$, making the SRE hard to compute for generic $n$.

\begin{figure*}
	\includegraphics[width=0.9\textwidth]{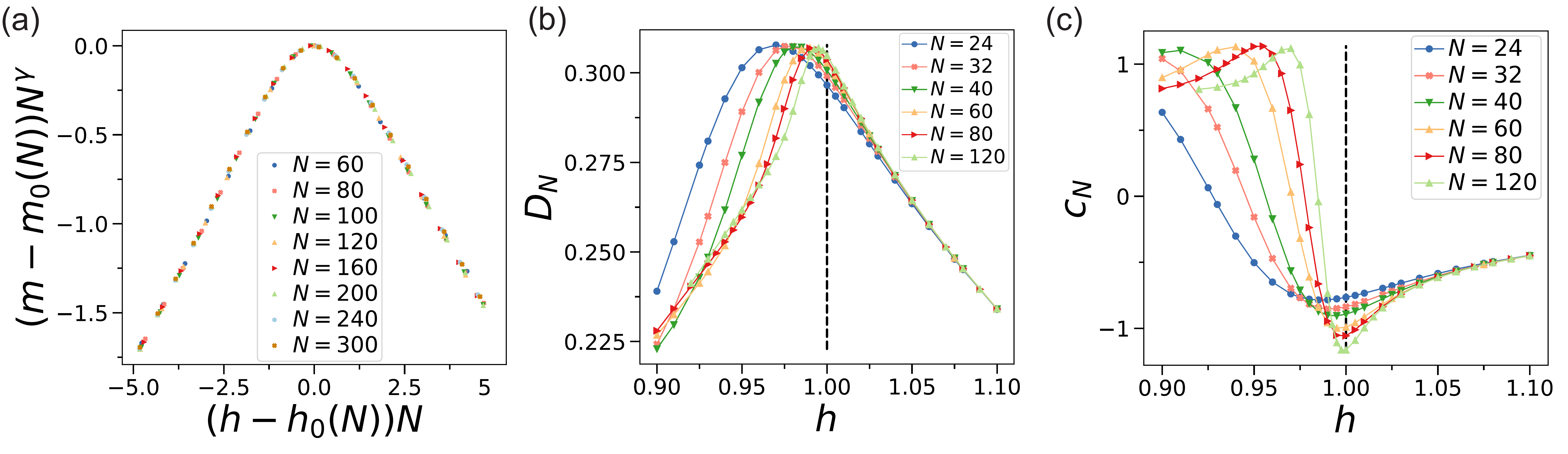}
	\caption{$(a)$: Scaling behavior of density of magic for increasing system sizes. $m_0(N)$ and $h_0(N)$ are, respectively, the maximum of $m(h,N)$ and the value of $h$ for which $m(h,N)$ is maximal. Here $\gamma\simeq 0.85$ is chosen in order to have the best data collapse. $(b)$, $(c)$: Linear coefficient and sublinear terms, defined via $M^{(2)}(\ket{\Psi^{\rm gs}_N})=D_N N+c_N$. For each $N$, $D_N$ and $c_N$ are extracted following the procedure explained in the main text with $\delta N=4$. For large $N$, $D_N$ coincides with the density $m$.}
	\label{fig:renyi_entropies}
\end{figure*}

\subsection{Computability from the replica MPS}
As our first result, we show that the SRE can be computed efficiently for integer $n>1$. The idea  is based on the simple identity
\begin{align}\label{eq:contraction}
	 \sum_{P \in \mathcal{P}_{N}} \frac{\braket{\Psi_N|P|\Psi_N}^{2n}}{2^N}&=
	(\bra{\Psi_N}\otimes \bra{\Psi^\ast_N})^{\otimes n}\Lambda^{(n)}_1\otimes \Lambda^{(n)}_2\otimes\nonumber\\
	 &\cdots \otimes \Lambda^{(n)}_N (\ket{\Psi_N}\otimes \ket{\Psi^\ast_N})^{\otimes n}\,,
\end{align}
where $\Lambda^{(n)}_j=(1/2)\sum_{\alpha=0}^3 (\sigma^\alpha_j\otimes\sigma^{\alpha\ast}_j )^{\otimes n}$, while $(\cdot)^\ast$ denotes complex conjugation. Eq.~\eqref{eq:contraction} can be verified by elementary algebraic manipulations. The r.h.s. of Eq.~\eqref{eq:contraction} can be interpreted as the norm of a ``$2n$-replica'' MPS $\ket{\Phi^{(n)}_{N}}$, with bond dimension $\chi^\prime=\chi^{2n}$ and physical dimension $d^{(n)}=2^{2(n-1)}$. To see this, we first note that $ (\ket{\Psi_N}\otimes \ket{\Psi^\ast_N})^{\otimes n}$ is an MPS with bond dimension $\chi^{2n}$ and physical dimension $2^{2n}$. Next, it is easy to verify that $\Lambda^{(n)}_k\geq 0$ and ${\rm rank}(\Lambda^{(n)}_k)=2^{2(n-1)}$. Therefore, we can write $\Lambda^{(n)}_k=\Gamma^\dagger_k \Gamma_k$, and define the new tensor $B_k=\Gamma_k \cdot (A_k\otimes A^\ast_k)^{\otimes n}$. Thus, we have 
\begin{equation}\label{eq:norm_replica}
\frac{1}{2^N}\sum_{P\in\mathcal{P}_N}\braket{\Psi_N|P|\Psi_N}^{2n}= \braket{\Phi^{(n)}_N|\Phi^{(n)}_N},
\end{equation}
with 
\begin{equation}\label{eq:replica_MPS}
\!\! \ket{\Phi^{(n)}_N}\!\!=\!\!\sum_{\{\tilde{s}_{k}\}} {\rm tr}\left([S\otimes S^\ast]^{\otimes n} B^{\tilde{s}_{1}}_1 \ldots B^{\tilde{s}_{N}}_N\right)\!\!\ket{\tilde{s}_{1}, \ldots,\tilde{s}_{N}},
\end{equation}
where now $\tilde{s}_k=0,\ldots ,d^{(n)}-1$.
\begin{align}\label{eq:final_form}
\!\!M^{(n)}(|\Psi_N\rangle)=(1-n)^{-1}[\ln \braket{\Phi^{(n)}_N|\Phi^{(n)}_N}-\ln \mathcal{N}^{2n}]\,,
\end{align}
where $\mathcal{N}=\braket{\Psi_N|\Psi_N}$. This replica approach is reminiscent of a similar trick used in the study of the so-called participation entropy~\cite{luitz2014participation,stephan2009shannon,stephan2009renyi,alcaraz2013,stephan2014renyi}. In this context, it has served as the basis of both analytical methods~\cite{fradkin2006entanglement,hsu2010universal} and Monte Carlo numerical calculations~\cite{luitz2014participation,luitz2014universal,luitz2014improving}.

We note that one can derive an alternative expression for~\eqref{eq:contraction}, without complex conjugation. To this end, we observe that, since $P$ is Hermitian, we have both $|\braket{\Psi_N|P|\Psi_N}|^2=\braket{\Psi_N|P|\Psi_N}\braket{\Psi^\ast_N|P^\ast|\Psi^\ast_N}$ and $|\braket{\Psi_N|P|\Psi_N}|^2=\braket{\Psi_N|P|\Psi_N}^2$. Using the latter we arrive at an expression similar to~\eqref{eq:contraction} where $\Lambda_j^{(n)}$ is replaced by $(1/2)\sum_{\alpha=0}^3 (\sigma^\alpha_j)^{\otimes 2n}$. Note, however, that this is not a positive operator for $n$ odd, so that in this case we can not proceed to write a relation such as~\eqref{eq:norm_replica}. We will make use of the alternative expression for $n=2$ later, cf. Eq.~\eqref{eq:n2_contraction}.

Eq.~\eqref{eq:final_form} has important ramifications, as we first illustrate for TI MPS~\footnote{The same discussion holds for MPSs which are invariant under shift of $p$ sites, with $p>1$.}. In this case, $B_j=B$, independent of $j$. Introducing the \emph{transfer matrix}~\cite{cirac2017matrix_op}
\begin{equation}
\tau=\sum_{\tilde{s}=0}^{d^{(n)}-1} B^{\tilde{s}}\otimes B^{\tilde{s}\ast},
\end{equation}
and recalling $S=\openone$, we have $\braket{\Phi^{(n)}_N|\Phi^{(n)}_N}={\rm tr}(\tau^N)=\sum_k \lambda_k^{N}$. Here $\{\lambda_k\}$ is the set of (complex) eigenvalues of $\tau$. This result is interesting: it states that magic, a global quantity, is completely determined by the spectrum of a single local tensor, $\tau$, whose dimensions do not scale with $N$. In fact, this construction allows us to study directly the thermodynamic limit $N\to\infty$. Assuming $\tau$ has a single largest eigenvalue $\lambda^{(n)}_0$~\footnote{This is a working hypothesis encoding ``typical behavior'' of MPSs, and which simplifies our derivations. However, we do not expect it to be necessary, see also Appendix~\ref{sec:locality}.}, and that the state is normalized in the thermodynamic limit, we have
\begin{equation}
m^{(n)}:=\lim_{N\to\infty} M^{(n)}(\ket{\Psi_N})/N=(1-n)^{-1}\ln\lambda^{(n)}_0.
\end{equation}
Magic is thus extensive, and the asymptotic value of its density is a function of the leading eigenvalue of $\tau$.

We can make a step further, showing that $m^{(n)}$ can be computed \emph{locally}. To this end, consider a region $A$ of $\ell$ qubits and assume $N\gg \ell$. Denoting by $\rho_{A}$ the reduced density matrix on $A$, we introduce a local probe for the density of SRE 
\begin{equation}
m^{(n)}_\ell=-\frac{1}{\ell}\ln\frac{1}{2^{\ell}}\sum_{P\in\mathcal{P}_\ell}\left({\rm tr }[\rho_A P]\right)^{2n}
\end{equation}
We note that this differs from the formula for the R\'enyi-$2$ stabilizer entropy of mixed states in~\cite{leone2021quantum}, and here is intended as a local probe of pure-state magic. However, since MPSs satisfy an entanglement area law, the two definitions give the same density for large $\ell$. Using that $\tau$ has a single largest eigenvalue $\lambda^{(n)}_0$, we show in Appendix~\ref{sec:locality} that $m^{(n)}_\ell=m^{(n)}+O(1/\ell)$, \emph{i.e.} $m^{(n)}$ can be extracted from measuring a finite region of $\ell$ sites, up to an error $O(1/\ell)$. This result generalizes a similar observation made in~\cite{oliviero2022magic} for the GS of the quantum Ising model to TI MPSs, putting it on rigorous grounds. 

When $S=\ket{R}\bra{L}$, the state $\ket{\Phi_N^{(n)}}$ is an MPS with OBCs, and we may assume $\mathcal{N}=1$. Its norm can be computed exactly at a cost~\cite{schollwock2011density} $O(Nd^{(n)}\chi^{\prime 3})=O(N2^{2(n-1)}\chi^{6n})$, which is \emph{linear} in $N$, as previously announced. From the practical point of view, the bottleneck for numerical computations comes from the factor $\chi^{6n}$. However, for $n=2$ one can exploit additional symmetries, further reducing the computational cost. 

To see this, note that, for $n=2$, the r.h.s. of Eq.~\eqref{eq:contraction} can be rewritten as 
\begin{equation}\label{eq:n2_contraction}
\bra{\Psi_N}^{\otimes 4}\Lambda^{(2)}_1\otimes\cdots \otimes \Lambda^{(2)}_N \ket{\Psi_N}^{\otimes 4}=\braket{\tilde{\Phi}^{(2)}_N|\tilde{\Phi}^{(2)}_N},
\end{equation}
and $\Lambda^{(2)}_j=(1/2)\sum_{\alpha=0}^3(\sigma^{\alpha})^{\otimes 4}$, so that no complex conjugation appears. Here $\ket{\tilde{\Phi}^{(2)}_N}$ is the MPS with OBCs generated by $\tilde{B}_k=\Gamma_k \cdot (A_k)^{\otimes 4}$, and with boundaries $\ket{R}^{\otimes 4}$, $\bra{L}^{\otimes 4}$. The tensors $\tilde{B}_k$ manifestly commute with the linear representation of the Klein four group $\mathcal{K}=\{\openone, S_{12}S_{34},S_{13}S_{24},S_{14}S_{23}\}$, where $S_{jk}$ is the SWAP operator exchanging replica spaces $j, k$. Therefore, the auxiliary space decomposes into irreducible representations of $\mathcal{K}$. In fact, because of OBCs, the only irreducible representation allowed is the trivial one. Projecting onto the corresponding subspace, we compress $\chi$ to $\tilde{\chi}=(1/4)\chi^2(3+\chi^2)$, reducing the computational cost, see Appendix~\ref{sec:compression_mps} for details. 

\section{SRE in the quantum Ising model} 
\label{sec:SRE_QIM}
We apply the previously developed MPS approach to study magic in the GS of the quantum Ising model (with OBCs)
\begin{equation}\label{eq:hamiltonian}
	H_\text{Ising}=-\sum_{k=1}^{N-1}\sigma_{k}^x\sigma_{k+1}^x-h\sum_{k=1}^{N}\sigma_{k}^z\,,
\end{equation}
where $h$ is a magnetic field. The model is exactly solvable via the Jordan-Wigner (JW) mapping, and displays a quantum phase transition at $h=h_c=1$~\cite{sachdev2011}.  GS magic of the quantum Ising chain was recently investigated in Refs.~\cite{sarkar2020characterization,oliviero2022magic}. While Ref.~\cite{sarkar2020characterization} focused on the one- and two-site GS reduced density matrix, Ref.~\cite{oliviero2022magic}  computed the stabilizer R\'enyi-$2$ entropy of the whole chain (with PBCs), based on its exact solution. The method, however, involved a computational cost exponential in $N$ and was limited to sizes up to $N=12$~\cite{oliviero2022magic}. We revisit the study of GS magic for the Hamiltonian~\eqref{eq:hamiltonian}, obtaining accurate numerical data up to $N\simeq 300$,  significantly extending previous analyses. 

Our approach is based on approximating the GS of~\eqref{eq:hamiltonian} as an MPS using the standard density-matrix renormalization group (DMRG) algorithm~\cite{schollwock2011density} implemented with the ITensor library~\cite{itensor}, and exploiting~\eqref{eq:final_form} to compute the stabilizer R\'enyi-$2$ entropy. 
Let us denote by $\ket{\Psi_N(\chi)}$ an MPS approximation for the true ground-state $\ket{\Psi_N^{\rm gs}}$, with bond dimension $\chi$. The efficiency of this method depends on how the difference $\Delta=|M^{(2)}(\ket{\Psi_N(\chi)})-M^{(2)}(\ket{\Psi_N^{\rm gs}})|$ scales with the fidelity $F=|\braket{\Psi_N(\chi)|\Psi_N^{\rm gs}}|^2$. Comparing against exact-diagonalization calculations up to $N=12$, we verified that, roughly, $\Delta\sim |1-F|^{0.5}$, so that $|1-F|$ is not required to be exponentially small in $N$. In practice, in all our computations, we always verified that our results are stable upon increasing $\chi$, and we see that relatively small bond dimensions are enough to approximate $M^{(2)}(\ket{\Psi_N^{\rm gs}})$ up to good accuracy. Further detail and additional numerical data are reported in Appendix~\ref{sec:additional_data}.

\begin{figure*}
	\includegraphics[width=0.9\textwidth]{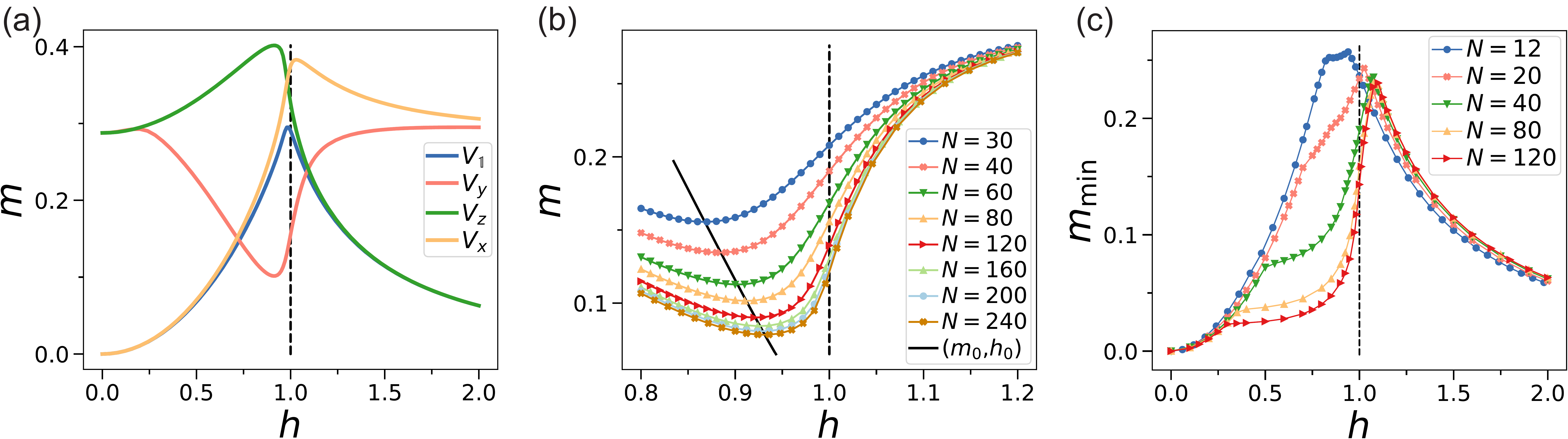}
	\caption{$(a)$: Density of magic $m$ against field $h$ for different bases defined by single-qubit rotations $V_{\alpha}=\exp(-i\frac{1}{2}\frac{\pi}{4}\sigma^\alpha)$, where $\sigma^\alpha\in\{\openone,\sigma^x,\sigma^y,\sigma^z\}$ are the Pauli operators. We show $N=80$ and $\chi=8$. $(b)$: $m$ for the rotated basis $V_y$ close to the critical point. Solid black
	line shows the fit of minimum magic $m_0=c_m N^{-\eta_m}+b_m$ and corresponding field $h_0=c_hN^{-\eta_h}+b_h$ with asymptotic limit $b_h=0.943(1)$, $b_m=0.065(1)$. We clearly see that the SRE is not maximal for $h=1$ (dashed line as guide to the eye). $(c)$: Minimal magic over the set of all local basis transformations. Numerical minimization is repeated 5 times to avoid local minima. By increasing $N$, $m$ can be reduced significantly for $h<1$, being maximal close to (but not at) $h=1$.}
	\label{fig:magic_2}
\end{figure*}

We studied the density 
\begin{equation}
m(h,N)=M^{(2)}(\ket{\Psi_N^{\rm gs}})/N
\end{equation}
as a function of $N$ for different values of $h$. \footnote{For $|h|<1$ the GS is two-fold degenerate for $N\to\infty$. Following Ref.~\cite{oliviero2022magic}, we focused on the exact GS at finite $N$, which is symmetric with respect to the $\mathbb{Z}_2$ symmetry $Z=\prod_j\sigma^z_j$, but we expect that the density of magic is the same for the two short-range correlated symmetry-broken GSs.}. For small system sizes, we recover the results of Ref.~\cite{oliviero2022magic}: away from $h=1$, we find $m\propto h^{2}$ for $h\ll1$ and $m\propto h^{-2}$ for $h\gg1$. In addition, $m(h,N)$ displays its maximum, denoted by $m_0(N)$ for a value $h_0(N)$ approaching $h_c=1$ for $N\to\infty$. The large system sizes available allow us to study the behavior near $h_c$ more closely. We have computed $m_0(N)$ and $h_0(N)$ up to $N=300$, and fitted both sets of data against the functions 
\begin{subequations}
\begin{align}
h_0(N)&=c_hN^{-\eta_h}+b_h\,,\\
m_0(N)&=c_m N^{-\eta_m}+b_m\,,
\end{align}
\end{subequations}
for the parameters $c_h$, $b_h$, $\eta_h$ and $c_m$, $b_m$, $\eta_m$. We find $b_h=0.9996(3)$, $\eta_h=1.078(18)$, and $b_m=0.3080(6)$, $\eta_m=0.665(34)$. Here we report the error associated with the fitting procedure including an estimation of the error due to finite bond dimension.

Next, motivated by the scaling of entanglement near criticality~\cite{osterloh2002scaling,calabrese2004entanglement,amico2008entanglement,calabrese2009entanglement}, we investigated the emergence of a universal scaling behavior, plotting the data against the rescaled variable $(h-h_0(N))N^{1/\nu}$, with $\nu=1$~\cite{osterloh2002scaling}. Fig.~\ref{fig:renyi_entropies}$(a)$ shows our results $m(h,N)$ shifted by its maximum $m_0(N)$. We observe excellent data collapse when rescaling $m(h,N)-m_0(N)$ by $N^\gamma$ with $\gamma\simeq 0.85$.

It is interesting to discuss the connections with the participation entropy, which has been extensively studied in one-dimensional systems~\cite{stephan2009shannon,stephan2009renyi,alcaraz2013,stephan2014renyi}. Similar to~\eqref{eq:SRE}, it is defined as the R\'enyi entropy of a classical probability distribution function $|\braket{\Psi^{\rm gs}_N|i_1,\ldots i_N}|^{2}$, where $\ket{i_j}$ is a local computational basis~\cite{luitz2014participation}. In the Ising model, it was shown to scale linearly in $N$, while its subleading $O(1)$ term displays a universal step-like profile as a function of $h$~\cite{stephan2009shannon}. Inspired by these studies, we define the linear coefficient $D_N$ and the sublinear term $c_N$ via 
\begin{equation}
M^{(2)}(\ket{\Psi^{\rm gs}_N})=:D_N N+c_N\,.
\end{equation}
In order to extract $D_N$, $c_N$, we exploit the procedure explained in Ref.~\cite{sierant2022universal}: we compute $M^{(2)}(\ket{\Psi^{\rm gs}_N})$ for sets of three sizes $N-\delta N$, $N$ and $N+\delta N$ with small $\delta N$ and fit the corresponding three values against the straight line $D_N N+c_N$. The result of our analysis is reported in Figs.~\ref{fig:renyi_entropies}$(b)$, $(c)$. Contrary to the participation entropy~\cite{stephan2009renyi}, $c_N$ does not display a universal step-like profile for the available system sizes. Still, it appears to develop a discontinuity at $h_c=1$. Overall, these findings confirm that different features of the SRE detect the presence of the quantum phase transition, substantiating the results presented in Ref.~\cite{leone2021quantum}. 

The definition of the SRE strongly depends on the computational basis, and an important question is whether some of the previously observed features are independent from it. Therefore, we have studied the SRE in different bases, obtained by acting on the system with $V^{\otimes N}$, where $V$ is a single-qubit unitary. We found that the behavior of the density $m(h,N)$ is not universal, in the sense that it is strongly basis dependent. In Fig.~\ref{fig:magic_2}$(a)$ we report data for different choices of $V$, while Fig.~\ref{fig:magic_2}$(b)$ shows data for a rotation of an angle $\theta=\pi/4$ around the $y$-axis, $V_y=\exp(-i\frac{1}{2}\theta\sigma^y)$. We clearly see that $V_y$ does not develop an extremum at $h_c=1$. This is confirmed by a fit $h=0.946(3)$. In light of this analysis, the unrotated basis appears to be special, as $m$ displays a maximum at criticality. This could be explained by the fact that the Hamiltonian is written precisely in terms of the Pauli matrices, although this point deserves further investigations. Finally, the coefficient $c_N$ appears to develop a discontinuity at the critical point, independent of the chosen basis, see Appendix~\ref{sec:additional_data} for additional numerical data. Therefore, the behavior of $c_N$ seemingly captures the phase transition, in analogy to the participation entropy~\cite{luitz2014participation}.

Overall, our findings suggest that a significant part of GS magic is short-ranged, even at criticality, as a large fraction of it can be removed by strictly local unitary transformations. In order to investigate this point further, we set up an optimization scheme to look for the local unitary transformation minimizing magic for a given value of $h$ and $N$. This can be done by a simple global optimisation approach~\cite{nelder1965simplex} in the space of single-qubit unitaries $V$. In Fig.~\ref{fig:magic_2}$(c)$, we find that the minimal density of magic $m_{\rm min}$ displays a clear peak close to $h_c=1$.
This analysis confirms the intuition that criticality is associated to long-range magic~\cite{white2021conformal,ellison2021symmetry}. Note that the peak of $m_{\rm min}$ in Eq.~\eqref{fig:magic_2} is not exactly at $h_c=1$. This could be due to the fact that local rotations are not the most general unitary transformations with a finite correlation length. We expect that performing an optimization over a larger family of local transformations, such as quantum circuits of increasing finite depth, will result in the maximum of $m_{\rm min}$ to approach $h_c$. 

\section{Outlook}\label{sec:outlook}
We developed a replica approach to study the SRE~\cite{leone2021quantum} of MPSs. In the TI case, we showed that the SRE can be expressed entirely in terms of the spectrum of a suitably defined transfer matrix, while it can be computed efficiently for MPSs with OBCs. We illustrated the usefulness of this construction by computing the R\'enyi-$2$ stabilizer entropy in the Ising chain, significantly expanding previous analyses~\cite{oliviero2022magic}. By investigating the dependence of the SRE on different local bases, we unveiled a more subtle connection between magic and criticality than previously expected. Our work opens up many directions. The method presented here could be straightforwardly applied to GSs of more general one-dimensional models, probing the role played by integrability and quantum chaos. In addition, our replica approach could be applied in different classes of Tensor-Network states such as PEPS~\cite{cirac2020matrix} or Tree-Tensor Networks~\cite{silvi2019tensor}, opening the way to investigate many-body quantum magic in higher dimensions.

\begin{acknowledgments}
\prlsection{Acknowledgments} We are grateful to Xhek Turkeshi for very useful discussions, and especially for drawing our attention to the participation entropy.
\end{acknowledgments}

\appendix

\section{Locality of magic for TI MPS}
\label{sec:locality}
In this Appendix, we provide further details on the SRE of TI MPSs. We show in particular that the SRE can be computed locally. To this end, we consider a region $A$ of $\ell$ qubits  and, denoting by $\rho_{A}$ the reduced density matrix on $A$. We define
\begin{equation}\label{eq:local_m}
m^{(n)}_\ell=-\lim_{N\to\infty}\frac{1}{\ell}\ln\frac{1}{2^{\ell}}\sum_{P\in\mathcal{P}_N}\left({\rm tr }[\rho_A P]\right)^{2n}\,.
\end{equation}
Our goal is to show that, for TI MPSs $\ket{\Psi_N}$, we have
\begin{equation}\label{eq:to_prove}
m^{(n)}_\ell=\lim_{N\to\infty}\frac{M^{(n)}(\ket{\Psi_N})}{N} +O(1/\ell)\,.
\end{equation}
We consider an MPS  $\ket{\Psi_N}$ with PBCs $\ket{\Psi_{N}}=\sum_{\{s_{k}\}} {\rm tr}\left(A^{s_{1}} \ldots A^{s_{N}}\right)\ket{s_{1}, \ldots, s_{N}}$ with transfer matrix
\begin{equation}\label{eq:tau_A}
	\tau(A)=\sum_{s=0}^{1} A^{s}\otimes A^{s\ast}\,.
\end{equation}
We assume that $\lim_{N\to\infty}\braket{\Psi_N|\Psi_N}=1$, without loss of generality. Next, we introduce the replica MPSs
\begin{subequations}\label{eq:replica_MPS_SM}
\begin{align}
	\ket{\Phi^{(n)}_N}&=\sum_{\{\tilde{s}_{k}\}} {\rm tr}\left(B^{\tilde{s}_{1}}_1 \ldots B^{\tilde{s}_{N}}_N\right)\ket{\tilde{s}_{1}, \ldots,\tilde{s}_{N}}\,,\\
 \ket{\Psi^{(n)}_N}&=\sum_{\{\tilde{s}_{k}\}} {\rm tr}\left(C^{\tilde{s}_{1}}_1 \ldots C^{\tilde{s}_{N}}_N\right)\ket{\tilde{s}_{1}, \ldots,\tilde{s}_{N}}\,,
\end{align}
\end{subequations}
where $B_k=\Gamma_k \cdot (A_k\otimes A^\ast_k)^{\otimes n}$, and $\Gamma_k$ is given by 
\begin{equation}
\Lambda^{(n)}_k=(1/2)\sum_{\alpha=0}^3 (\sigma^\alpha_k\otimes\sigma^{\alpha\ast}_k )^{\otimes n}=\Gamma^\dagger_k\Gamma_k,
\end{equation}
while $C_k=(A_k\otimes A^\ast_k)^{\otimes n}$. Finally, we define the corresponding transfer matrices
\begin{subequations}\label{eq:transfer_matrices}
\begin{align}
	\tau_{\Phi}^{(n)}(B)&=\sum_{\tilde{s}=0}^{d^{(n)}-1} B^{\tilde{s}}\otimes B^{\tilde{s}\ast}\,,\\ 
	\tau_{\Psi}^{(n)}(C)&=\sum_{\tilde{s}=0}^{2^{n}-1} C^{\tilde{s}}\otimes C^{\tilde{s}\ast}\,,
\end{align}
\end{subequations}
where $d^{(n)}=2^{(n-1)}$, cf. the main text. 

We assume that $A$ is normal~\cite{cirac2017matrix_op}, that is $(i)$ there exists no non-trivial projector $\Pi$ such that $A_i\Pi = \Pi A_i \Pi$; $(ii)$ the associated completely positive map (CPM) $\varepsilon_{A}(\cdot)=\sum_{i=1}^{d} A^{i} (\cdot )A^{i\dagger}$ has a unique eigenvalue of magnitude (and value) equal to its spectral radius, which is equal to one. Then, the transfer matrix $\tau(A)$ in \eqref{eq:tau_A} has unique left and right eigenvectors $\ket{R}$, $\bra{L}$, corresponding to eigenvalues $\lambda=1$ (and no other eigenvalue $\nu$ with $|\nu|=1$). Clearly, the same is true for $\tau_{\Psi}^{(n)}(C)$, with leading eigenstates $\ket{R^{(n)}_{\Psi}}:=(\ket{R}\otimes \ket{R^\ast})^{\otimes n}$, $\bra{L^{(n)}_{\Psi}}:=(\bra{L}\otimes \bra{L^\ast})^{\otimes n}$.

\begin{figure*}
	\includegraphics[scale=0.22]{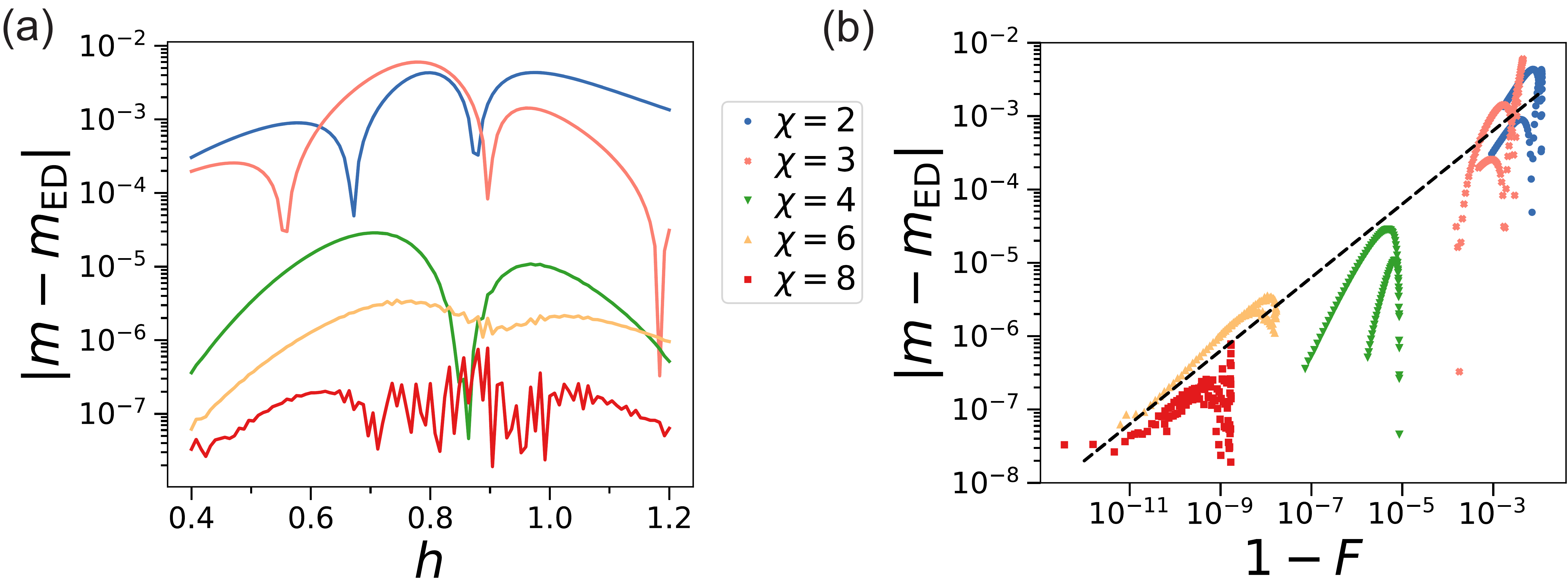}
	\caption{$(a)$: Error $\Delta=|m-m_\text{ED}|$ in magic density $m$ via MPS compared to exact diagonalization $M_\text{ED}$ for $N = 12$ and various $\chi$.  $(b)$: $\Delta$ plotted against fidelity $F$. Dashed line is a guide to the eye showing $\Delta\propto |1-F|^{0.5}$. }
	\label{fig:fidelity_1}
\end{figure*}

\begin{figure*}
	\includegraphics[scale=0.22]{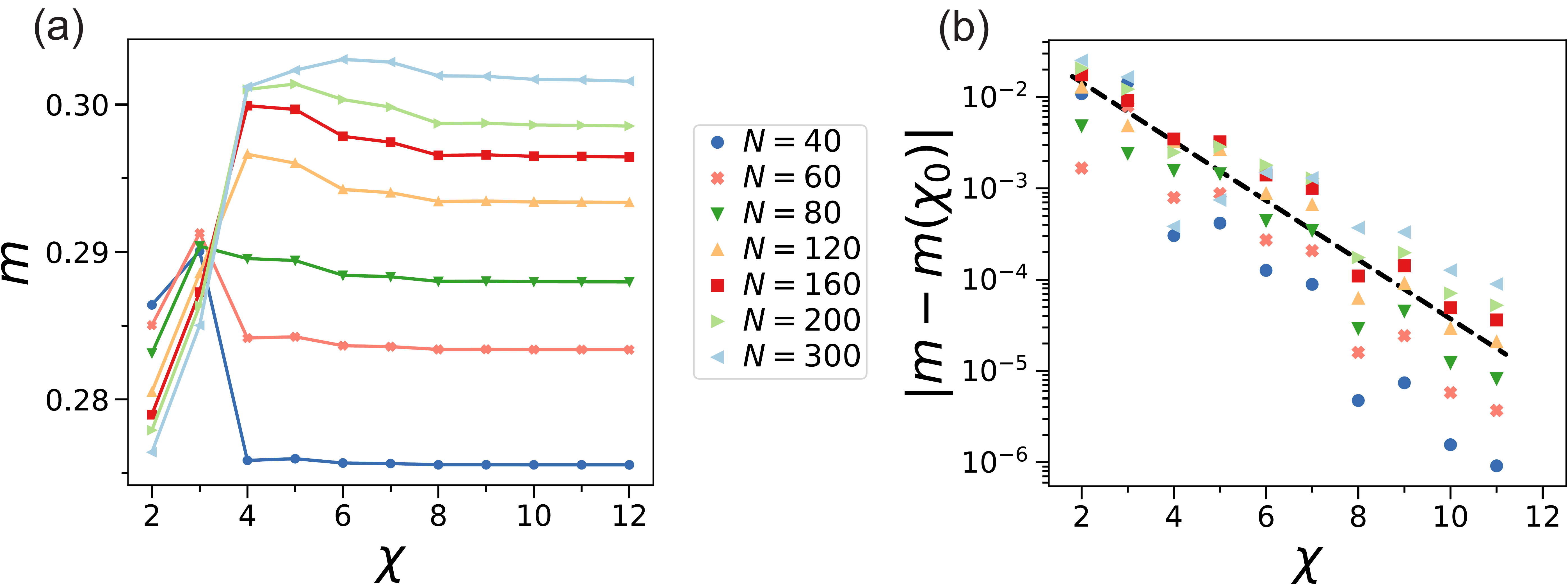}
	\caption{$(a)$: Density of SRE, $m$, of the GS as function of $\chi$ for various system sizes $N$ close to $h\approx1$. $(b)$: Difference of $m$ computed for bond dimension $\chi$ and  $\chi_0=12$. Dashed line is a fit with $\vert m-m(\chi_0)\vert\propto 10^{-\gamma\chi}$ with $\gamma=0.32$. }
	\label{fig:fidelity_2}
\end{figure*}

\begin{figure*}
	\includegraphics[width=0.75\textwidth]{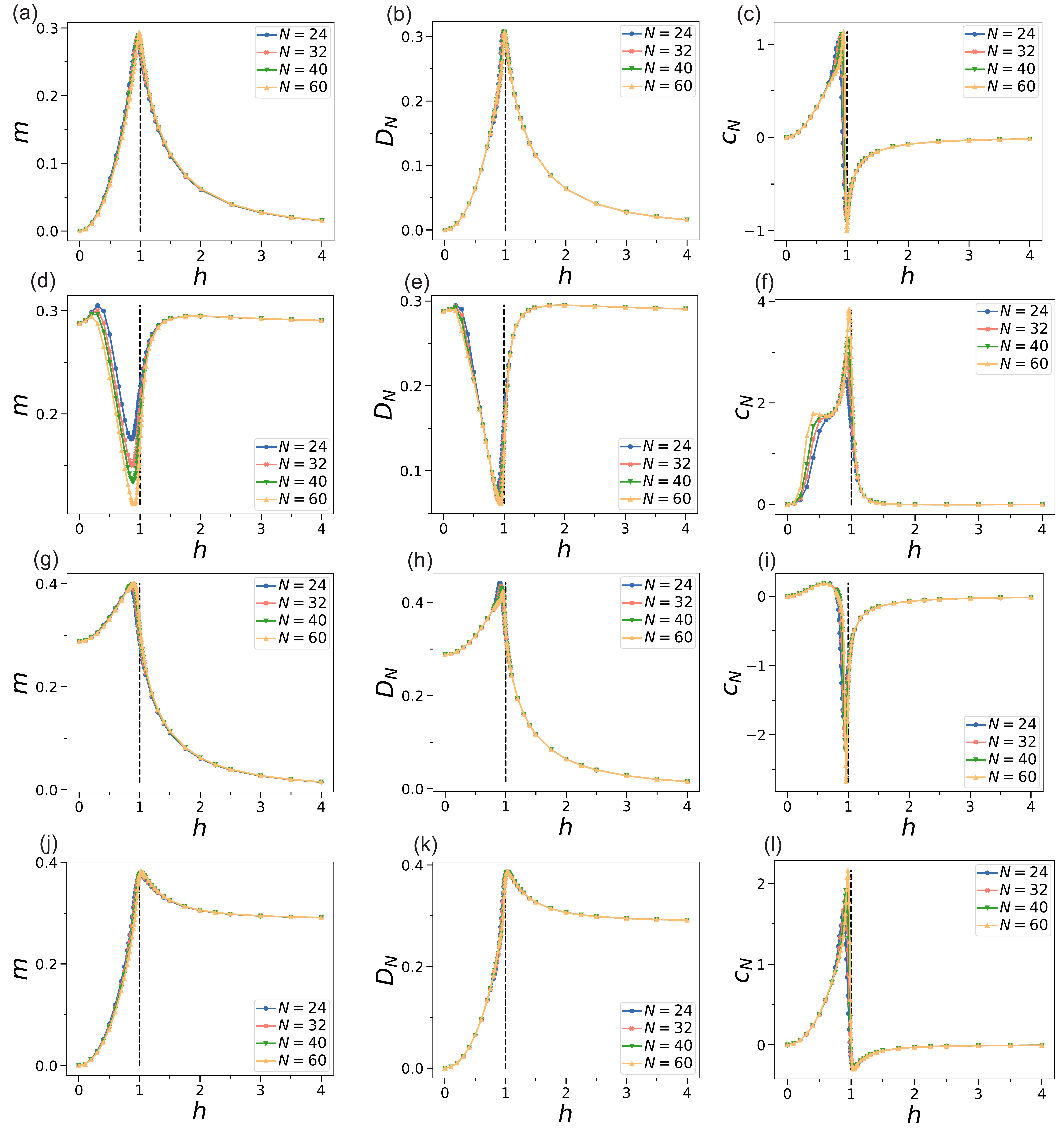}
	\caption{Magic density $m$, linear coefficient $D_N$ and sublinear term $c_N$ as defined via $M^{(2)}(\ket{\Psi^{\rm gs}_N})=D_N N+c_N$ for the GS of Ising model. Extraction procedure is explained in the main text. We show four different bases $V_{\alpha}=\exp(-i\frac{1}{2}\frac{\pi}{4}\sigma^\alpha)$, where $\sigma^\alpha\in\{\openone,\sigma^x,\sigma^y,\sigma^z\}$ are the Pauli operators. (a,d,g,h) shows $m$, $(b,e,h,k)$: $D_N$, and $(c,f,i,l)$: $c_N$. We show four bases, namely
	$(a,b,c)$: unrotated basis, $(d,e,f)$: $V_y$, $(g,h,i)$: $V_z$, and $(j,k,l)$: $V_x$.
	}
	\label{fig:fit_1}
\end{figure*}

\begin{figure*}
	\includegraphics[width=0.75\textwidth]{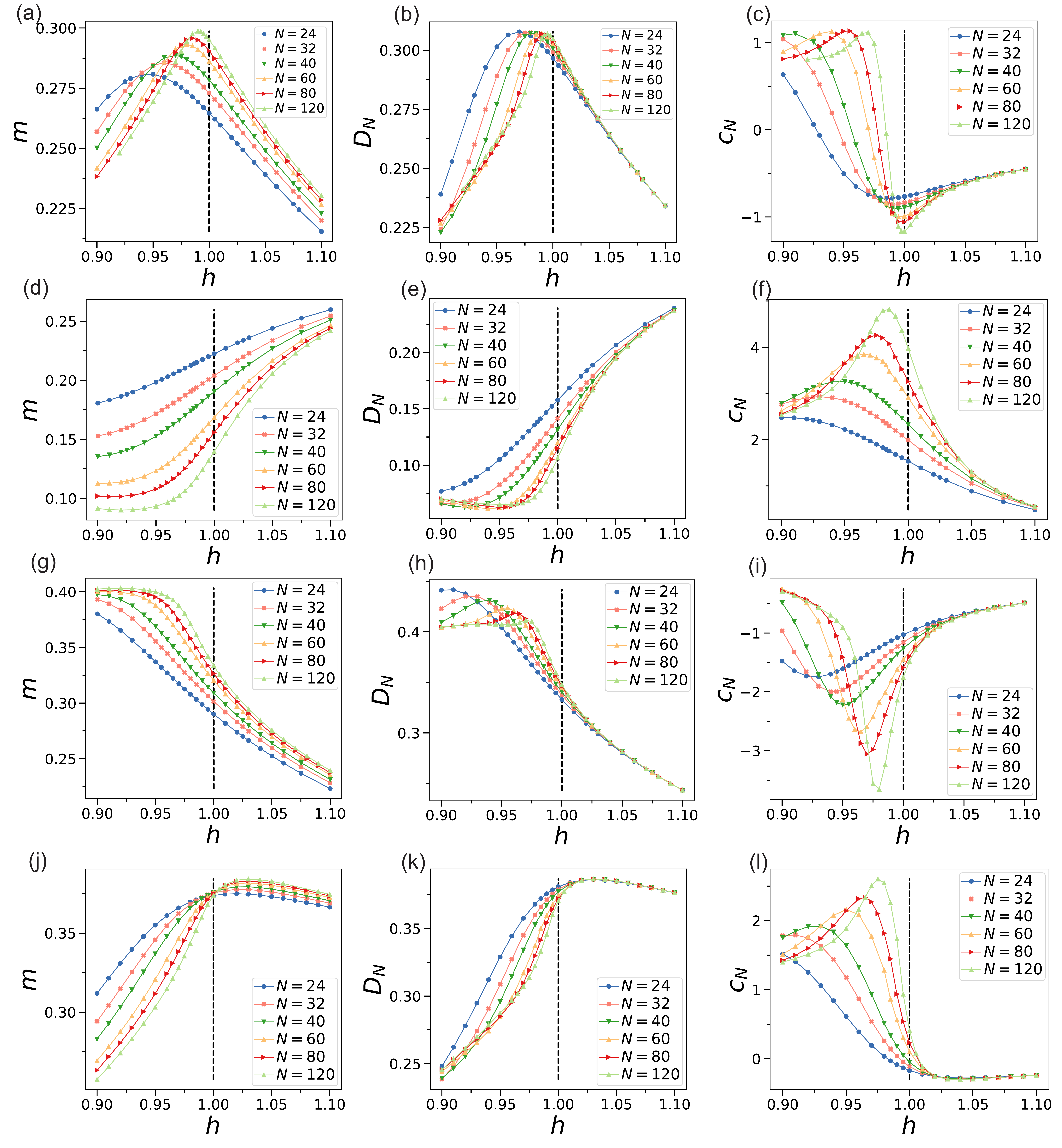}
	\caption{Magic density $m$, linear coefficient $D_N$ and sublinear term $c_N$ close to criticality. We show same parameters as in Fig.\ref{fig:fit_1} with $(a,d,g,j)$: $m$, $(b,e,h,k)$: $D_N$ and $(c,f,i,l)$: $c_N$. We show four bases with
	$(a,b,c)$: unrotated basis, $(d,e,f)$: $V_y$, $(g,h,i)$: $V_z$, and $(j,k,l)$: $V_x$. Magic is computed with bond dimension $\chi=8$ for $N\le80$ and $\chi=10$ else.
	}
	\label{fig:fit_2}
\end{figure*}

In addition, we also assume that $\tau_{\Phi}^{(n)}(B)$ has a unique eigenvalue of magnitude (and value) equal to its spectral radius, denoted by $\lambda^{(n)}_0$. Note that this condition alone does not imply that $B$ is normal. This is a working hypothesis encoding ``typical behavior'' of MPSs, and which simplifies our derivations. However, we do not expect it to be necessary. In fact, numerical evidence suggests that this condition always holds if $A$ is normal, although we were not able to prove it. We will denote the right/left eigenstates associated with $\lambda^{(n)}_0$ by $\ket{R^{(n)}_{\Phi}}$ and $\bra{L^{(n)}_{\Phi}}$. Note that
\begin{widetext}
\begin{equation}\label{eq:limit_m}
	\lim_{N\to\infty}\frac{M^{(n)}(\ket{\Psi_N})}{N}=\lim_{N\to\infty} (1-n)^{-1}\frac{1}{N}\ln {\rm tr}[\tau_{\Phi}^{(n)}(B)^{N}]=(1-n)^{-1}\ln\lambda^{(0)}_n\,.
\end{equation}
Let us now consider a chain of $N$ sites, and set $R=N-\ell$.  Using the same replica approach explained in the main text, we can rewrite the argument of the logarithm in~\eqref{eq:local_m} as
\begin{equation}
\frac{1}{2^{\ell}}\sum_{P\in\mathcal{P}_N}\left({\rm tr }[\rho_A P]\right)^{2n}=	\frac{1}{2^{\ell}}\sum_{\{\alpha_j\}_{j=1}^{\ell}}\braket{\Psi_N| \sigma^{\alpha_1}_1\ldots \sigma^{\alpha_{\ell}}_{\ell}\openone_{\ell+1}\cdots \openone_N|\Psi_N}^{2n}=	{\rm tr}\{[\tau_{\Phi}^{(n)}(B)]^{\ell}[\tau_{\Psi}^{(n)}(C)]^{R}\}\,.
\end{equation}
Taking the limit $R\to\infty$, we get
\begin{equation}
\lim_{R\to\infty}{\rm tr}\{[\tau_{\Phi}^{(n)}(B)]^{\ell}[\tau_{\Psi}^{(n)}(C)]^{R}\}=\braket{L^{(n)}_{\Psi}|\tau_{\Phi}^{(n)}(B)]^{\ell}|R^{(n)}_{\Psi}}= (\lambda^{(0)}_n)^{\ell}  \braket{L^{(n)}_{\Psi}| R^{(n)}_{\Phi}} \braket{L^{(n)}_{\Phi}| R^{(n)}_{\Psi}}(1+O(e^{-\alpha\ell}))\,,
\end{equation}
\end{widetext}
for some $\alpha>0$. In order to conclude, it is enough to show that $\braket{L^{(n)}_{\Psi}| R^{(n)}_{\Phi}}>0$ (and analogously for $\braket{L^{(n)}_{\Phi}| R^{(n)}_{\Psi}}$). To see that this is true, note first that $\braket{L^{(n)}_{\Psi}| R^{(n)}_{\Phi}}={\rm tr}\{L^{(n)}_{\Psi}R^{(n)}_{\Phi}\}$ where $L^{(n)}_{\Psi}$, $R^{(n)}_{\Phi}$ are the matrix representations of $\bra{L^{(n)}_{\Psi}}$, $\ket{ R^{(n)}_{\Phi}}$, \emph{i.e.} the eigenvectors of the CPMs $\varepsilon^\dagger_{C}$ and $\varepsilon_{B}$, respectively. Next, we use that since $A$ is normal,  $L^{(n)}_{\Psi}$ has full rank, and that both $L^{(n)}_{\Psi}$, $R^{(n)}_{\Phi}$ are positive operators. This follow from the fact that, for a CPM with a single non-degenerate eigenvalue on its peripheral spectrum, the corresponding eigenstate is a positive operator~\cite{evans1977spectral}.

\section{Compression of the replica MPS for R\'enyi-$2$ stabilizer entropy}
\label{sec:compression_mps}

We provide some technical details on the implementation of the replica MPS for R\'enyi-$2$ SRE.

Let $\ket{\Psi_N}$ be an MPS with OBCs. First, we note that, for $n=2$, the r.h.s. of Eq.3 in the main text can be rewritten as 
\begin{equation}
\bra{\Psi_N}^{\otimes 4}\Lambda^{(2)}_1\otimes\cdots \otimes \Lambda^{(2)}_N \ket{\Psi_N}^{\otimes 4}=\braket{\tilde{\Phi}^{(2)}_N|\tilde{\Phi}^{(2)}_N}
\end{equation}
with $\Lambda^{(2)}_j=(1/2)\sum_{\alpha=0}^3(\sigma^{\alpha})^{\otimes 4}$, so that no complex conjugation appears. Here $\ket{\tilde{\Phi}^{(2)}_N}$ is the replica MPS with OBCs generated by the tensors $\tilde{B}_k=\Gamma_k \cdot (A_k)^{\otimes 4}$, and with boundary vectors $\ket{R}^{\otimes 4}$, $\bra{L}^{\otimes 4}$. Using the definition of $\Gamma_k$, the local tensors $\tilde{B}_k$ now manifestly commute with the elements of the linear representation of the Klein four group 
\begin{equation}
\mathcal{K}=\{\openone, S_{12}S_{34},S_{13}S_{24},S_{14}S_{23}\}\,,
\end{equation}
where $S_{jk}$ is the SWAP operator exchanging replica spaces $j$ and $k$. Therefore, the auxiliary space decomposes into irreducible representations of $\mathcal{K}$. Because of OBCs, the only possible irreducible representation is the trivial one. Therefore, we may insert in the auxiliary space the projector
\begin{equation}\label{eq:proj}
	\Pi=\frac{1}{4}\left(\openone + S_{12}S_{34}+S_{13}S_{24}+S_{14}S_{23}\right)\,.
\end{equation}
It is easy to compute 
\begin{equation}
{\rm rank}(\Pi)=(1/4)\chi^2(3+\chi^2)\,.
\end{equation}
Then, we find a matrix $Q$ such that $\Pi=Q^\dagger Q$, and we can reduce the bond dimension by defining the new tensor
\begin{equation}
 C=Q\tilde{B} Q^{\dagger}\,.
\end{equation}
Importantly, $Q$ is sparse, and one can construct $C$ without ever constructing the full matrix $\tilde{B}$. 

To illustrate the construction, we consider the transformation of the left link of the local tensor, i.e. $C'=Q A^{\otimes 4}$. First, we note that we can construct $Q$ by considering the action of $\Pi$ on the set of basis states $\ket{a_1,a_2,a_3,a_4}$, with $a_i\in\{0,\dots,\chi-1\}$ and $i=1,\dots,4$, which represent the bond indices of $(A_k)^{\otimes 4}$. $\Pi$ applied to a basis state yields a linear combination of states, \emph{e.g.} $\Pi\ket{1200}\sim \ket{1200}+\ket{2100}+\ket{0012}+\ket{0021}$.  The rows of $Q$ can be written as all unique transformations of $\Pi\ket{a_1,a_2,a_3,a_4}$ including proper normalisation. Here, each row consists of only up to $4$ non-zero entries. Then, we construct $QA^{\otimes 4}$ by computing each entry of $Q$ individually, without performing explicit matrix multiplication. This turns out to be numerically faster and less memory consuming as we do not need to explicitly construct the full tensor $A^{\otimes 4}$.

\section{Additional numerical data}
\label{sec:additional_data}

In this Appendix we provide additional numerical data for the GS SRE in the Quantum Ising model.

\subsection{Accuracy and dependence with $\chi$}

We begin by studying the accuracy of our MPS-based approach. As mentioned in the main text, we have first compared it against exact-diagonalization (ED) data, which can be performed for small system sizes. In Fig.~\ref{fig:fidelity_1}, we plot the difference $\Delta=|m-m_\text{ED}|$ between the density of Renyi-$2$ stabilizer entropy in the GS of the Ising Hamiltonian, computed using ED ($m_\text{ED}$) and our method ($m$). Fig.~\ref{fig:fidelity_1}$(b)$ shows in particular the difference as a function of $h$ for increasing bond dimension $\chi$. For $N=12$, we see that the latter is very small ($~10^{-7}$) already for $\chi=6$. In Fig.~\ref{fig:fidelity_1}$(b)$ we investigate the dependence of $\Delta$ with the fidelity $F=|\braket{\Psi_N(\chi)|\Psi_N^{\rm gs}}|^2$. Different sets of points of the same color correspond to data produced for different values of $h$ and the same bond dimension $\chi$. The dashed line is a guide for the eye, showing that, roughly $\Delta\sim |1-F|^{0.5}$.

For larger system sizes, ED data are not available, but we have always tested that our data are well converged upon increasing the bond dimension. An example is shown in Fig.~\ref{fig:fidelity_2}$(a)$. In general, we see that, as $N$ increases, a larger bond dimension is needed in order to have faithful results. Still, the data appear to be converged already for $\chi=10$, up to $N=300$. In Fig.~\ref{fig:fidelity_2} we also plot the difference between the density of SRE of two MPS approximations, $\ket{\Psi_N(\chi)}$ and $\ket{\Psi_N(\chi_0=12)}$, for $\chi=2,..,11$. The plot shows a convincing exponential decay of the error as a function of $\chi$ for all sizes, further supporting the accuracy of the method.

\subsection{Additional data for rotated bases}

Finally, we provide additional data for the density of SRE in rotated bases. We consider in particular the linear coefficient  $D_N$ and subleading term $c_N$ defined via 
\begin{equation}
M^{(2)}(\ket{\Psi^{\rm gs}_N})=:D_N N+c_N.
\end{equation}
As mentioned in the main text, we have extracted them exploiting the procedure explained in Ref.~\cite{sierant2022universal}. Namely we computed $M^{(2)}(\ket{\Psi^{\rm gs}_N})$ for sets of three sizes $N-\delta N$, $N$ and $N+\delta N$ with small $\delta N$ and fit the corresponding three values against the straight line $D_N N+c_N$
We report in Figs.~\ref{fig:fit_1}, data for three different bases as a function of $h$.

Fig.~\ref{fig:fit_2} shows the same data close to criticality.
Clearly, $D_N$ coincides with the density of SRE for large-$N$. Consistently, while it displays a maximum for the unrotated basis, this is not the case in general. This is reported in~\ref{fig:fit_2}, showing that $D_N$ either displays a maximum or a minimum that is extremal away from $h_c$ for the chosen bases. Polynomial fits of the extremum with 
\begin{equation}
h_0(N)=c_hN^{-\eta_h}+b_h
\end{equation}
suggest that this effect persists even in the limit of large $N$.
In contrast to $m$ and $D_N$, we find that the coefficient $c_N$ becomes extremal close to $h_c$. The extremum can be fitted very well with 
\begin{equation}
c_{N,0}=a_c\ln(N)+b_c,
\end{equation}
suggesting that it is logarithmically diverging. A polynomial fit of the extremal field $h^{c_N}_0$ for $c_N$ is consistent with $\lim_{N\to\infty}h^{c_N}_0= h_c=1$, at least within the numerical accuracy. These findings persist for the four types of bases $V_\alpha$ we investigated, suggesting that $c_N$ is able to diagnose the phase transition, independent of the local basis.

\bibliography{bibliography}

\begin{thebibliography}{70}%
\makeatletter
\providecommand \@ifxundefined [1]{%
 \@ifx{#1\undefined}
}%
\providecommand \@ifnum [1]{%
 \ifnum #1\expandafter \@firstoftwo
 \else \expandafter \@secondoftwo
 \fi
}%
\providecommand \@ifx [1]{%
 \ifx #1\expandafter \@firstoftwo
 \else \expandafter \@secondoftwo
 \fi
}%
\providecommand \natexlab [1]{#1}%
\providecommand \enquote  [1]{``#1''}%
\providecommand \bibnamefont  [1]{#1}%
\providecommand \bibfnamefont [1]{#1}%
\providecommand \citenamefont [1]{#1}%
\providecommand \href@noop [0]{\@secondoftwo}%
\providecommand \href [0]{\begingroup \@sanitize@url \@href}%
\providecommand \@href[1]{\@@startlink{#1}\@@href}%
\providecommand \@@href[1]{\endgroup#1\@@endlink}%
\providecommand \@sanitize@url [0]{\catcode `\\12\catcode `\$12\catcode
  `\&12\catcode `\#12\catcode `\^12\catcode `\_12\catcode `\%12\relax}%
\providecommand \@@startlink[1]{}%
\providecommand \@@endlink[0]{}%
\providecommand \url  [0]{\begingroup\@sanitize@url \@url }%
\providecommand \@url [1]{\endgroup\@href {#1}{\urlprefix }}%
\providecommand \urlprefix  [0]{URL }%
\providecommand \Eprint [0]{\href }%
\providecommand \doibase [0]{http://dx.doi.org/}%
\providecommand \selectlanguage [0]{\@gobble}%
\providecommand \bibinfo  [0]{\@secondoftwo}%
\providecommand \bibfield  [0]{\@secondoftwo}%
\providecommand \translation [1]{[#1]}%
\providecommand \BibitemOpen [0]{}%
\providecommand \bibitemStop [0]{}%
\providecommand \bibitemNoStop [0]{.\EOS\space}%
\providecommand \EOS [0]{\spacefactor3000\relax}%
\providecommand \BibitemShut  [1]{\csname bibitem#1\endcsname}%
\let\auto@bib@innerbib\@empty
\bibitem [{\citenamefont {Feynman}(1982)}]{feynman1982simulating}%
  \BibitemOpen
  \bibfield  {author} {\bibinfo {author} {\bibfnamefont {R.~P.}\ \bibnamefont
  {Feynman}},\ }\href {\doibase 10.1007/BF02650179} {\bibfield  {journal}
  {\bibinfo  {journal} {Int. J. Theor. Phys.}\ }\textbf {\bibinfo {volume}
  {21}},\ \bibinfo {pages} {467} (\bibinfo {year} {1982})}\BibitemShut
  {NoStop}%
\bibitem [{\citenamefont {Gottesman}(1997)}]{gottesman1997stabilizer}%
  \BibitemOpen
  \bibfield  {author} {\bibinfo {author} {\bibfnamefont {D.}~\bibnamefont
  {Gottesman}},\ }\emph {\bibinfo {title} {Stabilizer codes and quantum error
  correction. Caltech Ph. D}},\ \href@noop {} {Ph.D. thesis},\ \bibinfo
  {school} {Thesis, eprint: quant-ph/9705052} (\bibinfo {year}
  {1997})\BibitemShut {NoStop}%
\bibitem [{\citenamefont
  {Gottesman}(1998{\natexlab{a}})}]{gottesman1998theory}%
  \BibitemOpen
  \bibfield  {author} {\bibinfo {author} {\bibfnamefont {D.}~\bibnamefont
  {Gottesman}},\ }\href {\doibase 10.1103/PhysRevA.57.127} {\bibfield
  {journal} {\bibinfo  {journal} {Phys. Rev. A}\ }\textbf {\bibinfo {volume}
  {57}},\ \bibinfo {pages} {127} (\bibinfo {year}
  {1998}{\natexlab{a}})}\BibitemShut {NoStop}%
\bibitem [{\citenamefont
  {Gottesman}(1998{\natexlab{b}})}]{gottesman1998heisenberg}%
  \BibitemOpen
  \bibfield  {author} {\bibinfo {author} {\bibfnamefont {D.}~\bibnamefont
  {Gottesman}},\ }\href {https://arxiv.org/abs/quant-ph/9807006} {\bibfield
  {journal} {\bibinfo  {journal} {arXiv quant-ph/9807006}\ } (\bibinfo {year}
  {1998}{\natexlab{b}})}\BibitemShut {NoStop}%
\bibitem [{\citenamefont {Aaronson}\ and\ \citenamefont
  {Gottesman}(2004)}]{aaronson2004improved}%
  \BibitemOpen
  \bibfield  {author} {\bibinfo {author} {\bibfnamefont {S.}~\bibnamefont
  {Aaronson}}\ and\ \bibinfo {author} {\bibfnamefont {D.}~\bibnamefont
  {Gottesman}},\ }\href {\doibase 10.1103/PhysRevA.70.052328} {\bibfield
  {journal} {\bibinfo  {journal} {Phys. Rev. A}\ }\textbf {\bibinfo {volume}
  {70}},\ \bibinfo {pages} {052328} (\bibinfo {year} {2004})}\BibitemShut
  {NoStop}%
\bibitem [{\citenamefont {Nielsen}\ and\ \citenamefont
  {Chuang}(2011)}]{nielsen2011quantum}%
  \BibitemOpen
  \bibfield  {author} {\bibinfo {author} {\bibfnamefont {M.~A.}\ \bibnamefont
  {Nielsen}}\ and\ \bibinfo {author} {\bibfnamefont {I.~L.}\ \bibnamefont
  {Chuang}},\ }\href@noop {} {\emph {\bibinfo {title} {Quantum Computation and
  Quantum Information: 10th Anniversary Edition}}}\ (\bibinfo  {publisher}
  {Cambridge University Press},\ \bibinfo {year} {2011})\BibitemShut {NoStop}%
\bibitem [{\citenamefont {Shor}(1996)}]{shor1996fault}%
  \BibitemOpen
  \bibfield  {author} {\bibinfo {author} {\bibfnamefont {P.~W.}\ \bibnamefont
  {Shor}},\ }in\ \href@noop {} {\emph {\bibinfo {booktitle} {Proceedings of
  37th conference on foundations of computer science}}}\ (\bibinfo
  {organization} {IEEE},\ \bibinfo {year} {1996})\ pp.\ \bibinfo {pages}
  {56--65}\BibitemShut {NoStop}%
\bibitem [{\citenamefont {Preskill}(1998)}]{preskill1998fault}%
  \BibitemOpen
  \bibfield  {author} {\bibinfo {author} {\bibfnamefont {J.}~\bibnamefont
  {Preskill}},\ }in\ \href@noop {} {\emph {\bibinfo {booktitle} {Introduction
  to quantum computation and information}}}\ (\bibinfo  {publisher} {World
  Scientific},\ \bibinfo {year} {1998})\ pp.\ \bibinfo {pages}
  {213--269}\BibitemShut {NoStop}%
\bibitem [{\citenamefont {Kitaev}(2003)}]{kitaev2003fault}%
  \BibitemOpen
  \bibfield  {author} {\bibinfo {author} {\bibfnamefont {A.~Y.}\ \bibnamefont
  {Kitaev}},\ }\href {\doibase 10.1016/S0003-4916(02)00018-0} {\bibfield
  {journal} {\bibinfo  {journal} {Ann. Phys.}\ }\textbf {\bibinfo {volume}
  {303}},\ \bibinfo {pages} {2} (\bibinfo {year} {2003})}\BibitemShut {NoStop}%
\bibitem [{\citenamefont {Eastin}\ and\ \citenamefont
  {Knill}(2009)}]{eastin2009restriction}%
  \BibitemOpen
  \bibfield  {author} {\bibinfo {author} {\bibfnamefont {B.}~\bibnamefont
  {Eastin}}\ and\ \bibinfo {author} {\bibfnamefont {E.}~\bibnamefont {Knill}},\
  }\href {\doibase 10.1103/PhysRevLett.102.110502} {\bibfield  {journal}
  {\bibinfo  {journal} {Phys. Rev. Lett.}\ }\textbf {\bibinfo {volume} {102}},\
  \bibinfo {pages} {110502} (\bibinfo {year} {2009})}\BibitemShut {NoStop}%
\bibitem [{\citenamefont {Bravyi}\ and\ \citenamefont
  {Kitaev}(2005)}]{bravyi2005universal}%
  \BibitemOpen
  \bibfield  {author} {\bibinfo {author} {\bibfnamefont {S.}~\bibnamefont
  {Bravyi}}\ and\ \bibinfo {author} {\bibfnamefont {A.}~\bibnamefont
  {Kitaev}},\ }\href {\doibase 10.1103/PhysRevA.71.022316} {\bibfield
  {journal} {\bibinfo  {journal} {Phys. Rev. A}\ }\textbf {\bibinfo {volume}
  {71}},\ \bibinfo {pages} {022316} (\bibinfo {year} {2005})}\BibitemShut
  {NoStop}%
\bibitem [{\citenamefont {Campbell}\ \emph {et~al.}(2017)\citenamefont
  {Campbell}, \citenamefont {Terhal},\ and\ \citenamefont
  {Vuillot}}]{campbell2017roads}%
  \BibitemOpen
  \bibfield  {author} {\bibinfo {author} {\bibfnamefont {E.~T.}\ \bibnamefont
  {Campbell}}, \bibinfo {author} {\bibfnamefont {B.~M.}\ \bibnamefont
  {Terhal}}, \ and\ \bibinfo {author} {\bibfnamefont {C.}~\bibnamefont
  {Vuillot}},\ }\href {\doibase 10.1038/nature23460} {\bibfield  {journal}
  {\bibinfo  {journal} {Nature}\ }\textbf {\bibinfo {volume} {549}},\ \bibinfo
  {pages} {172} (\bibinfo {year} {2017})}\BibitemShut {NoStop}%
\bibitem [{\citenamefont {Nahum}\ \emph {et~al.}(2017)\citenamefont {Nahum},
  \citenamefont {Ruhman}, \citenamefont {Vijay},\ and\ \citenamefont
  {Haah}}]{nahum2017quantum}%
  \BibitemOpen
  \bibfield  {author} {\bibinfo {author} {\bibfnamefont {A.}~\bibnamefont
  {Nahum}}, \bibinfo {author} {\bibfnamefont {J.}~\bibnamefont {Ruhman}},
  \bibinfo {author} {\bibfnamefont {S.}~\bibnamefont {Vijay}}, \ and\ \bibinfo
  {author} {\bibfnamefont {J.}~\bibnamefont {Haah}},\ }\href {\doibase
  10.1103/PhysRevX.7.031016} {\bibfield  {journal} {\bibinfo  {journal} {Phys.
  Rev. X}\ }\textbf {\bibinfo {volume} {7}},\ \bibinfo {pages} {031016}
  (\bibinfo {year} {2017})}\BibitemShut {NoStop}%
\bibitem [{\citenamefont {Pastawski}\ \emph {et~al.}(2015)\citenamefont
  {Pastawski}, \citenamefont {Yoshida}, \citenamefont {Harlow},\ and\
  \citenamefont {Preskill}}]{pastawski2015holographic}%
  \BibitemOpen
  \bibfield  {author} {\bibinfo {author} {\bibfnamefont {F.}~\bibnamefont
  {Pastawski}}, \bibinfo {author} {\bibfnamefont {B.}~\bibnamefont {Yoshida}},
  \bibinfo {author} {\bibfnamefont {D.}~\bibnamefont {Harlow}}, \ and\ \bibinfo
  {author} {\bibfnamefont {J.}~\bibnamefont {Preskill}},\ }\href {\doibase
  10.1007/JHEP06(2015)149} {\bibfield  {journal} {\bibinfo  {journal} {JHEP}\
  }\textbf {\bibinfo {volume} {2015}},\ \bibinfo {pages} {1} (\bibinfo {year}
  {2015})}\BibitemShut {NoStop}%
\bibitem [{\citenamefont {Howard}\ \emph {et~al.}(2014)\citenamefont {Howard},
  \citenamefont {Wallman}, \citenamefont {Veitch},\ and\ \citenamefont
  {Emerson}}]{howard2014contextuality}%
  \BibitemOpen
  \bibfield  {author} {\bibinfo {author} {\bibfnamefont {M.}~\bibnamefont
  {Howard}}, \bibinfo {author} {\bibfnamefont {J.}~\bibnamefont {Wallman}},
  \bibinfo {author} {\bibfnamefont {V.}~\bibnamefont {Veitch}}, \ and\ \bibinfo
  {author} {\bibfnamefont {J.}~\bibnamefont {Emerson}},\ }\href {\doibase
  10.1038/nature13460} {\bibfield  {journal} {\bibinfo  {journal} {Nature}\
  }\textbf {\bibinfo {volume} {510}},\ \bibinfo {pages} {351} (\bibinfo {year}
  {2014})}\BibitemShut {NoStop}%
\bibitem [{\citenamefont {Bravyi}\ \emph {et~al.}(2016)\citenamefont {Bravyi},
  \citenamefont {Smith},\ and\ \citenamefont {Smolin}}]{bravyi2016trading}%
  \BibitemOpen
  \bibfield  {author} {\bibinfo {author} {\bibfnamefont {S.}~\bibnamefont
  {Bravyi}}, \bibinfo {author} {\bibfnamefont {G.}~\bibnamefont {Smith}}, \
  and\ \bibinfo {author} {\bibfnamefont {J.~A.}\ \bibnamefont {Smolin}},\
  }\href {\doibase 10.1103/PhysRevX.6.021043} {\bibfield  {journal} {\bibinfo
  {journal} {Phys. Rev. X}\ }\textbf {\bibinfo {volume} {6}},\ \bibinfo {pages}
  {021043} (\bibinfo {year} {2016})}\BibitemShut {NoStop}%
\bibitem [{\citenamefont {Bravyi}\ \emph {et~al.}(2019)\citenamefont {Bravyi},
  \citenamefont {Browne}, \citenamefont {Calpin}, \citenamefont {Campbell},
  \citenamefont {Gosset},\ and\ \citenamefont {Howard}}]{bravyi2019simulation}%
  \BibitemOpen
  \bibfield  {author} {\bibinfo {author} {\bibfnamefont {S.}~\bibnamefont
  {Bravyi}}, \bibinfo {author} {\bibfnamefont {D.}~\bibnamefont {Browne}},
  \bibinfo {author} {\bibfnamefont {P.}~\bibnamefont {Calpin}}, \bibinfo
  {author} {\bibfnamefont {E.}~\bibnamefont {Campbell}}, \bibinfo {author}
  {\bibfnamefont {D.}~\bibnamefont {Gosset}}, \ and\ \bibinfo {author}
  {\bibfnamefont {M.}~\bibnamefont {Howard}},\ }\href {\doibase
  10.22331/q-2019-09-02-181} {\bibfield  {journal} {\bibinfo  {journal}
  {Quantum}\ }\textbf {\bibinfo {volume} {3}},\ \bibinfo {pages} {181}
  (\bibinfo {year} {2019})}\BibitemShut {NoStop}%
\bibitem [{\citenamefont {Seddon}\ \emph {et~al.}(2021)\citenamefont {Seddon},
  \citenamefont {Regula}, \citenamefont {Pashayan}, \citenamefont {Ouyang},\
  and\ \citenamefont {Campbell}}]{seddon2021quantifying}%
  \BibitemOpen
  \bibfield  {author} {\bibinfo {author} {\bibfnamefont {J.~R.}\ \bibnamefont
  {Seddon}}, \bibinfo {author} {\bibfnamefont {B.}~\bibnamefont {Regula}},
  \bibinfo {author} {\bibfnamefont {H.}~\bibnamefont {Pashayan}}, \bibinfo
  {author} {\bibfnamefont {Y.}~\bibnamefont {Ouyang}}, \ and\ \bibinfo {author}
  {\bibfnamefont {E.~T.}\ \bibnamefont {Campbell}},\ }\href {\doibase
  10.1103/PRXQuantum.2.010345} {\bibfield  {journal} {\bibinfo  {journal} {PRX
  Quantum}\ }\textbf {\bibinfo {volume} {2}},\ \bibinfo {pages} {010345}
  (\bibinfo {year} {2021})}\BibitemShut {NoStop}%
\bibitem [{\citenamefont {Koukoulekidis}\ \emph {et~al.}(2022)\citenamefont
  {Koukoulekidis}, \citenamefont {Kwon}, \citenamefont {Jee}, \citenamefont
  {Jennings},\ and\ \citenamefont {Kim}}]{koukoulekidis2022born}%
  \BibitemOpen
  \bibfield  {author} {\bibinfo {author} {\bibfnamefont {N.}~\bibnamefont
  {Koukoulekidis}}, \bibinfo {author} {\bibfnamefont {H.}~\bibnamefont {Kwon}},
  \bibinfo {author} {\bibfnamefont {H.~H.}\ \bibnamefont {Jee}}, \bibinfo
  {author} {\bibfnamefont {D.}~\bibnamefont {Jennings}}, \ and\ \bibinfo
  {author} {\bibfnamefont {M.}~\bibnamefont {Kim}},\ }\href
  {https://arxiv.org/abs/2202.12114} {\bibfield  {journal} {\bibinfo  {journal}
  {arXiv:2202.12114}\ } (\bibinfo {year} {2022})}\BibitemShut {NoStop}%
\bibitem [{\citenamefont {Leone}\ \emph {et~al.}(2022)\citenamefont {Leone},
  \citenamefont {Oliviero},\ and\ \citenamefont {Hamma}}]{leone2022stabilizer}%
  \BibitemOpen
  \bibfield  {author} {\bibinfo {author} {\bibfnamefont {L.}~\bibnamefont
  {Leone}}, \bibinfo {author} {\bibfnamefont {S.~F.~E.}\ \bibnamefont
  {Oliviero}}, \ and\ \bibinfo {author} {\bibfnamefont {A.}~\bibnamefont
  {Hamma}},\ }\href {\doibase 10.1103/PhysRevLett.128.050402} {\bibfield
  {journal} {\bibinfo  {journal} {Phys. Rev. Lett.}\ }\textbf {\bibinfo
  {volume} {128}},\ \bibinfo {pages} {050402} (\bibinfo {year}
  {2022})}\BibitemShut {NoStop}%
\bibitem [{\citenamefont {Leone}\ \emph {et~al.}(2021)\citenamefont {Leone},
  \citenamefont {Oliviero}, \citenamefont {Zhou},\ and\ \citenamefont
  {Hamma}}]{leone2021quantum}%
  \BibitemOpen
  \bibfield  {author} {\bibinfo {author} {\bibfnamefont {L.}~\bibnamefont
  {Leone}}, \bibinfo {author} {\bibfnamefont {S.~F.}\ \bibnamefont {Oliviero}},
  \bibinfo {author} {\bibfnamefont {Y.}~\bibnamefont {Zhou}}, \ and\ \bibinfo
  {author} {\bibfnamefont {A.}~\bibnamefont {Hamma}},\ }\href
  {https://arxiv.org/abs/2102.08406} {\bibfield  {journal} {\bibinfo  {journal}
  {Quantum}\ }\textbf {\bibinfo {volume} {5}},\ \bibinfo {pages} {453}
  (\bibinfo {year} {2021})}\BibitemShut {NoStop}%
\bibitem [{\citenamefont {Haferkamp}(2022)}]{haferkamp2022random}%
  \BibitemOpen
  \bibfield  {author} {\bibinfo {author} {\bibfnamefont {J.}~\bibnamefont
  {Haferkamp}},\ }\href {https://arxiv.org/abs/2203.16571} {\bibfield
  {journal} {\bibinfo  {journal} {arXiv:2203.16571}\ } (\bibinfo {year}
  {2022})}\BibitemShut {NoStop}%
\bibitem [{\citenamefont {White}\ \emph {et~al.}(2021)\citenamefont {White},
  \citenamefont {Cao},\ and\ \citenamefont {Swingle}}]{white2021conformal}%
  \BibitemOpen
  \bibfield  {author} {\bibinfo {author} {\bibfnamefont {C.~D.}\ \bibnamefont
  {White}}, \bibinfo {author} {\bibfnamefont {C.}~\bibnamefont {Cao}}, \ and\
  \bibinfo {author} {\bibfnamefont {B.}~\bibnamefont {Swingle}},\ }\href
  {\doibase 10.1103/PhysRevB.103.075145} {\bibfield  {journal} {\bibinfo
  {journal} {Phys. Rev. B}\ }\textbf {\bibinfo {volume} {103}},\ \bibinfo
  {pages} {075145} (\bibinfo {year} {2021})}\BibitemShut {NoStop}%
\bibitem [{\citenamefont {Sarkar}\ \emph {et~al.}(2020)\citenamefont {Sarkar},
  \citenamefont {Mukhopadhyay},\ and\ \citenamefont
  {Bayat}}]{sarkar2020characterization}%
  \BibitemOpen
  \bibfield  {author} {\bibinfo {author} {\bibfnamefont {S.}~\bibnamefont
  {Sarkar}}, \bibinfo {author} {\bibfnamefont {C.}~\bibnamefont
  {Mukhopadhyay}}, \ and\ \bibinfo {author} {\bibfnamefont {A.}~\bibnamefont
  {Bayat}},\ }\href {\doibase 10.1088/1367-2630/aba919} {\bibfield  {journal}
  {\bibinfo  {journal} {New J. Phys.}\ }\textbf {\bibinfo {volume} {22}},\
  \bibinfo {pages} {083077} (\bibinfo {year} {2020})}\BibitemShut {NoStop}%
\bibitem [{\citenamefont {Sewell}\ and\ \citenamefont
  {White}(2022)}]{sewell2022mana}%
  \BibitemOpen
  \bibfield  {author} {\bibinfo {author} {\bibfnamefont {T.~J.}\ \bibnamefont
  {Sewell}}\ and\ \bibinfo {author} {\bibfnamefont {C.~D.}\ \bibnamefont
  {White}},\ }\href@noop {} {\bibfield  {journal} {\bibinfo  {journal}
  {Physical Review B}\ }\textbf {\bibinfo {volume} {106}},\ \bibinfo {pages}
  {125130} (\bibinfo {year} {2022})}\BibitemShut {NoStop}%
\bibitem [{\citenamefont {Oliviero}\ \emph
  {et~al.}(2022{\natexlab{a}})\citenamefont {Oliviero}, \citenamefont {Leone},\
  and\ \citenamefont {Hamma}}]{oliviero2022magic}%
  \BibitemOpen
  \bibfield  {author} {\bibinfo {author} {\bibfnamefont {S.~F.~E.}\
  \bibnamefont {Oliviero}}, \bibinfo {author} {\bibfnamefont {L.}~\bibnamefont
  {Leone}}, \ and\ \bibinfo {author} {\bibfnamefont {A.}~\bibnamefont
  {Hamma}},\ }\href {\doibase 10.1103/PhysRevA.106.042426} {\bibfield
  {journal} {\bibinfo  {journal} {Phys. Rev. A}\ }\textbf {\bibinfo {volume}
  {106}},\ \bibinfo {pages} {042426} (\bibinfo {year}
  {2022}{\natexlab{a}})}\BibitemShut {NoStop}%
\bibitem [{\citenamefont {Liu}\ and\ \citenamefont
  {Winter}(2022)}]{liu2022many}%
  \BibitemOpen
  \bibfield  {author} {\bibinfo {author} {\bibfnamefont {Z.-W.}\ \bibnamefont
  {Liu}}\ and\ \bibinfo {author} {\bibfnamefont {A.}~\bibnamefont {Winter}},\
  }\href {\doibase 10.1103/PRXQuantum.3.020333} {\bibfield  {journal} {\bibinfo
   {journal} {PRX Quantum}\ }\textbf {\bibinfo {volume} {3}},\ \bibinfo {pages}
  {020333} (\bibinfo {year} {2022})}\BibitemShut {NoStop}%
\bibitem [{\citenamefont {Campbell}(2011)}]{campbell2011catalysis}%
  \BibitemOpen
  \bibfield  {author} {\bibinfo {author} {\bibfnamefont {E.~T.}\ \bibnamefont
  {Campbell}},\ }\href {\doibase 10.1103/PhysRevA.83.032317} {\bibfield
  {journal} {\bibinfo  {journal} {Phys. Rev. A}\ }\textbf {\bibinfo {volume}
  {83}},\ \bibinfo {pages} {032317} (\bibinfo {year} {2011})}\BibitemShut
  {NoStop}%
\bibitem [{\citenamefont {Veitch}\ \emph {et~al.}(2014)\citenamefont {Veitch},
  \citenamefont {Mousavian}, \citenamefont {Gottesman},\ and\ \citenamefont
  {Emerson}}]{veitch2014resource}%
  \BibitemOpen
  \bibfield  {author} {\bibinfo {author} {\bibfnamefont {V.}~\bibnamefont
  {Veitch}}, \bibinfo {author} {\bibfnamefont {S.~H.}\ \bibnamefont
  {Mousavian}}, \bibinfo {author} {\bibfnamefont {D.}~\bibnamefont
  {Gottesman}}, \ and\ \bibinfo {author} {\bibfnamefont {J.}~\bibnamefont
  {Emerson}},\ }\href {https://doi.org/10.1088/1367-2630/16/1/013009}
  {\bibfield  {journal} {\bibinfo  {journal} {New Journal of Physics}\ }\textbf
  {\bibinfo {volume} {16}},\ \bibinfo {pages} {013009} (\bibinfo {year}
  {2014})}\BibitemShut {NoStop}%
\bibitem [{\citenamefont {Howard}\ and\ \citenamefont
  {Campbell}(2017)}]{howard2017application}%
  \BibitemOpen
  \bibfield  {author} {\bibinfo {author} {\bibfnamefont {M.}~\bibnamefont
  {Howard}}\ and\ \bibinfo {author} {\bibfnamefont {E.}~\bibnamefont
  {Campbell}},\ }\href {\doibase 10.1103/PhysRevLett.118.090501} {\bibfield
  {journal} {\bibinfo  {journal} {Phys. Rev. Lett.}\ }\textbf {\bibinfo
  {volume} {118}},\ \bibinfo {pages} {090501} (\bibinfo {year}
  {2017})}\BibitemShut {NoStop}%
\bibitem [{\citenamefont {Wang}\ \emph {et~al.}(2019)\citenamefont {Wang},
  \citenamefont {Wilde},\ and\ \citenamefont {Su}}]{wang2019quantifying}%
  \BibitemOpen
  \bibfield  {author} {\bibinfo {author} {\bibfnamefont {X.}~\bibnamefont
  {Wang}}, \bibinfo {author} {\bibfnamefont {M.~M.}\ \bibnamefont {Wilde}}, \
  and\ \bibinfo {author} {\bibfnamefont {Y.}~\bibnamefont {Su}},\ }\href
  {https://doi.org/10.1088/1367-2630/ab451d} {\bibfield  {journal} {\bibinfo
  {journal} {New Journal of Physics}\ }\textbf {\bibinfo {volume} {21}},\
  \bibinfo {pages} {103002} (\bibinfo {year} {2019})}\BibitemShut {NoStop}%
\bibitem [{\citenamefont {Beverland}\ \emph {et~al.}(2020)\citenamefont
  {Beverland}, \citenamefont {Campbell}, \citenamefont {Howard},\ and\
  \citenamefont {Kliuchnikov}}]{beverland2020lower}%
  \BibitemOpen
  \bibfield  {author} {\bibinfo {author} {\bibfnamefont {M.}~\bibnamefont
  {Beverland}}, \bibinfo {author} {\bibfnamefont {E.}~\bibnamefont {Campbell}},
  \bibinfo {author} {\bibfnamefont {M.}~\bibnamefont {Howard}}, \ and\ \bibinfo
  {author} {\bibfnamefont {V.}~\bibnamefont {Kliuchnikov}},\ }\href {\doibase
  10.1088/2058-9565/ab8963} {\bibfield  {journal} {\bibinfo  {journal} {Quantum
  Science and Technology}\ }\textbf {\bibinfo {volume} {5}},\ \bibinfo {pages}
  {035009} (\bibinfo {year} {2020})}\BibitemShut {NoStop}%
\bibitem [{\citenamefont {Jiang}\ and\ \citenamefont
  {Wang}(2021)}]{jiang2021lower}%
  \BibitemOpen
  \bibfield  {author} {\bibinfo {author} {\bibfnamefont {J.}~\bibnamefont
  {Jiang}}\ and\ \bibinfo {author} {\bibfnamefont {X.}~\bibnamefont {Wang}},\
  }\href {\doibase 10.1088/2058-9565/ab8963} {\bibfield  {journal} {\bibinfo
  {journal} {arXiv:2103.09999}\ } (\bibinfo {year} {2021}),\
  10.1088/2058-9565/ab8963}\BibitemShut {NoStop}%
\bibitem [{\citenamefont {Hahn}\ \emph {et~al.}(2022)\citenamefont {Hahn},
  \citenamefont {Ferraro}, \citenamefont {Hultquist}, \citenamefont {Ferrini},\
  and\ \citenamefont {Garc\'{\i}a-\'Alvarez}}]{hahn2022quantifying}%
  \BibitemOpen
  \bibfield  {author} {\bibinfo {author} {\bibfnamefont {O.}~\bibnamefont
  {Hahn}}, \bibinfo {author} {\bibfnamefont {A.}~\bibnamefont {Ferraro}},
  \bibinfo {author} {\bibfnamefont {L.}~\bibnamefont {Hultquist}}, \bibinfo
  {author} {\bibfnamefont {G.}~\bibnamefont {Ferrini}}, \ and\ \bibinfo
  {author} {\bibfnamefont {L.}~\bibnamefont {Garc\'{\i}a-\'Alvarez}},\ }\href
  {\doibase 10.1103/PhysRevLett.128.210502} {\bibfield  {journal} {\bibinfo
  {journal} {Phys. Rev. Lett.}\ }\textbf {\bibinfo {volume} {128}},\ \bibinfo
  {pages} {210502} (\bibinfo {year} {2022})}\BibitemShut {NoStop}%
\bibitem [{\citenamefont {Bu}\ \emph {et~al.}(2022)\citenamefont {Bu},
  \citenamefont {Garcia}, \citenamefont {Jaffe}, \citenamefont {Koh},\ and\
  \citenamefont {Li}}]{bu2022complexity}%
  \BibitemOpen
  \bibfield  {author} {\bibinfo {author} {\bibfnamefont {K.}~\bibnamefont
  {Bu}}, \bibinfo {author} {\bibfnamefont {R.~J.}\ \bibnamefont {Garcia}},
  \bibinfo {author} {\bibfnamefont {A.}~\bibnamefont {Jaffe}}, \bibinfo
  {author} {\bibfnamefont {D.~E.}\ \bibnamefont {Koh}}, \ and\ \bibinfo
  {author} {\bibfnamefont {L.}~\bibnamefont {Li}},\ }\href
  {https://arxiv.org/abs/2204.12051} {\bibfield  {journal} {\bibinfo  {journal}
  {arXiv:2204.12051}\ } (\bibinfo {year} {2022})}\BibitemShut {NoStop}%
\bibitem [{\citenamefont {Campbell}\ \emph {et~al.}(2012)\citenamefont
  {Campbell}, \citenamefont {Anwar},\ and\ \citenamefont
  {Browne}}]{campbell2012magic}%
  \BibitemOpen
  \bibfield  {author} {\bibinfo {author} {\bibfnamefont {E.~T.}\ \bibnamefont
  {Campbell}}, \bibinfo {author} {\bibfnamefont {H.}~\bibnamefont {Anwar}}, \
  and\ \bibinfo {author} {\bibfnamefont {D.~E.}\ \bibnamefont {Browne}},\
  }\href {\doibase 10.1103/PhysRevX.2.041021} {\bibfield  {journal} {\bibinfo
  {journal} {Phys. Rev. X}\ }\textbf {\bibinfo {volume} {2}},\ \bibinfo {pages}
  {041021} (\bibinfo {year} {2012})}\BibitemShut {NoStop}%
\bibitem [{\citenamefont {Anwar}\ \emph {et~al.}(2014)\citenamefont {Anwar},
  \citenamefont {Brown}, \citenamefont {Campbell},\ and\ \citenamefont
  {Browne}}]{anwar2014fast}%
  \BibitemOpen
  \bibfield  {author} {\bibinfo {author} {\bibfnamefont {H.}~\bibnamefont
  {Anwar}}, \bibinfo {author} {\bibfnamefont {B.~J.}\ \bibnamefont {Brown}},
  \bibinfo {author} {\bibfnamefont {E.~T.}\ \bibnamefont {Campbell}}, \ and\
  \bibinfo {author} {\bibfnamefont {D.~E.}\ \bibnamefont {Browne}},\ }\href
  {\doibase 10.1088/1367-2630/16/6/063038} {\bibfield  {journal} {\bibinfo
  {journal} {New J. Phys.}\ }\textbf {\bibinfo {volume} {16}},\ \bibinfo
  {pages} {063038} (\bibinfo {year} {2014})}\BibitemShut {NoStop}%
\bibitem [{\citenamefont {Campbell}(2014)}]{campbell2014enhanced}%
  \BibitemOpen
  \bibfield  {author} {\bibinfo {author} {\bibfnamefont {E.~T.}\ \bibnamefont
  {Campbell}},\ }\href {\doibase 10.1103/PhysRevLett.113.230501} {\bibfield
  {journal} {\bibinfo  {journal} {Phys. Rev. Lett.}\ }\textbf {\bibinfo
  {volume} {113}},\ \bibinfo {pages} {230501} (\bibinfo {year}
  {2014})}\BibitemShut {NoStop}%
\bibitem [{\citenamefont {Haug}\ and\ \citenamefont
  {Kim}(2023)}]{haug2022scalable}%
  \BibitemOpen
  \bibfield  {author} {\bibinfo {author} {\bibfnamefont {T.}~\bibnamefont
  {Haug}}\ and\ \bibinfo {author} {\bibfnamefont {M.}~\bibnamefont {Kim}},\
  }\href {\doibase 10.1103/PRXQuantum.4.010301} {\bibfield  {journal} {\bibinfo
   {journal} {PRX Quantum}\ }\textbf {\bibinfo {volume} {4}},\ \bibinfo {pages}
  {010301} (\bibinfo {year} {2023})}\BibitemShut {NoStop}%
\bibitem [{\citenamefont {Oliviero}\ \emph
  {et~al.}(2022{\natexlab{b}})\citenamefont {Oliviero}, \citenamefont {Leone},
  \citenamefont {Hamma},\ and\ \citenamefont {Lloyd}}]{oliviero2022measuring}%
  \BibitemOpen
  \bibfield  {author} {\bibinfo {author} {\bibfnamefont {S.~F.}\ \bibnamefont
  {Oliviero}}, \bibinfo {author} {\bibfnamefont {L.}~\bibnamefont {Leone}},
  \bibinfo {author} {\bibfnamefont {A.}~\bibnamefont {Hamma}}, \ and\ \bibinfo
  {author} {\bibfnamefont {S.}~\bibnamefont {Lloyd}},\ }\href
  {https://arxiv.org/abs/2204.00015} {\bibfield  {journal} {\bibinfo  {journal}
  {arXiv:2204.00015}\ } (\bibinfo {year} {2022}{\natexlab{b}})}\BibitemShut
  {NoStop}%
\bibitem [{\citenamefont {Perez-Garcia}\ \emph {et~al.}(2007)\citenamefont
  {Perez-Garcia}, \citenamefont {Verstraete}, \citenamefont {Wolf},\ and\
  \citenamefont {Cirac}}]{perez2007matrix}%
  \BibitemOpen
  \bibfield  {author} {\bibinfo {author} {\bibfnamefont {D.}~\bibnamefont
  {Perez-Garcia}}, \bibinfo {author} {\bibfnamefont {F.}~\bibnamefont
  {Verstraete}}, \bibinfo {author} {\bibfnamefont {M.}~\bibnamefont {Wolf}}, \
  and\ \bibinfo {author} {\bibfnamefont {J.}~\bibnamefont {Cirac}},\ }\href
  {https://arxiv.org/abs/quant-ph/0608197} {\bibfield  {journal} {\bibinfo
  {journal} {Quantum Inf. Comp.}\ }\textbf {\bibinfo {volume} {7}},\ \bibinfo
  {pages} {401} (\bibinfo {year} {2007})}\BibitemShut {NoStop}%
\bibitem [{\citenamefont {Cirac}\ \emph {et~al.}(2017)\citenamefont {Cirac},
  \citenamefont {Perez-Garcia}, \citenamefont {Schuch},\ and\ \citenamefont
  {Verstraete}}]{cirac2017matrix_op}%
  \BibitemOpen
  \bibfield  {author} {\bibinfo {author} {\bibfnamefont {J.~I.}\ \bibnamefont
  {Cirac}}, \bibinfo {author} {\bibfnamefont {D.}~\bibnamefont {Perez-Garcia}},
  \bibinfo {author} {\bibfnamefont {N.}~\bibnamefont {Schuch}}, \ and\ \bibinfo
  {author} {\bibfnamefont {F.}~\bibnamefont {Verstraete}},\ }\href {\doibase
  10.1016/j.aop.2016.12.030} {\bibfield  {journal} {\bibinfo  {journal} {Ann.
  Phys.}\ }\textbf {\bibinfo {volume} {378}},\ \bibinfo {pages} {100} (\bibinfo
  {year} {2017})}\BibitemShut {NoStop}%
\bibitem [{\citenamefont {Cirac}\ \emph {et~al.}(2021)\citenamefont {Cirac},
  \citenamefont {Perez-Garcia}, \citenamefont {Schuch},\ and\ \citenamefont
  {Verstraete}}]{cirac2020matrix}%
  \BibitemOpen
  \bibfield  {author} {\bibinfo {author} {\bibfnamefont {J.~I.}\ \bibnamefont
  {Cirac}}, \bibinfo {author} {\bibfnamefont {D.}~\bibnamefont {Perez-Garcia}},
  \bibinfo {author} {\bibfnamefont {N.}~\bibnamefont {Schuch}}, \ and\ \bibinfo
  {author} {\bibfnamefont {F.}~\bibnamefont {Verstraete}},\ }\href@noop {}
  {\bibfield  {journal} {\bibinfo  {journal} {Rev. Mod. Phys.}\ }\textbf
  {\bibinfo {volume} {93}},\ \bibinfo {pages} {045003} (\bibinfo {year}
  {2021})}\BibitemShut {NoStop}%
\bibitem [{\citenamefont {Or{\'u}s}(2014)}]{orus2014practical}%
  \BibitemOpen
  \bibfield  {author} {\bibinfo {author} {\bibfnamefont {R.}~\bibnamefont
  {Or{\'u}s}},\ }\href {\doibase 10.1016/j.aop.2014.06.013} {\bibfield
  {journal} {\bibinfo  {journal} {Annals of physics}\ }\textbf {\bibinfo
  {volume} {349}},\ \bibinfo {pages} {117} (\bibinfo {year}
  {2014})}\BibitemShut {NoStop}%
\bibitem [{\citenamefont {Verstraete}\ and\ \citenamefont
  {Cirac}(2006)}]{verstraete2006matrix}%
  \BibitemOpen
  \bibfield  {author} {\bibinfo {author} {\bibfnamefont {F.}~\bibnamefont
  {Verstraete}}\ and\ \bibinfo {author} {\bibfnamefont {J.~I.}\ \bibnamefont
  {Cirac}},\ }\href {\doibase 10.1103/PhysRevB.73.094423} {\bibfield  {journal}
  {\bibinfo  {journal} {Phys. Rev. B}\ }\textbf {\bibinfo {volume} {73}},\
  \bibinfo {pages} {094423} (\bibinfo {year} {2006})}\BibitemShut {NoStop}%
\bibitem [{\citenamefont {Schuch}\ \emph {et~al.}(2008)\citenamefont {Schuch},
  \citenamefont {Wolf}, \citenamefont {Verstraete},\ and\ \citenamefont
  {Cirac}}]{schuch2008entropy}%
  \BibitemOpen
  \bibfield  {author} {\bibinfo {author} {\bibfnamefont {N.}~\bibnamefont
  {Schuch}}, \bibinfo {author} {\bibfnamefont {M.~M.}\ \bibnamefont {Wolf}},
  \bibinfo {author} {\bibfnamefont {F.}~\bibnamefont {Verstraete}}, \ and\
  \bibinfo {author} {\bibfnamefont {J.~I.}\ \bibnamefont {Cirac}},\ }\href
  {\doibase 10.1103/PhysRevLett.100.030504} {\bibfield  {journal} {\bibinfo
  {journal} {Phys. Rev. Lett.}\ }\textbf {\bibinfo {volume} {100}},\ \bibinfo
  {pages} {030504} (\bibinfo {year} {2008})}\BibitemShut {NoStop}%
\bibitem [{\citenamefont {Schollw{\"o}ck}(2011)}]{schollwock2011density}%
  \BibitemOpen
  \bibfield  {author} {\bibinfo {author} {\bibfnamefont {U.}~\bibnamefont
  {Schollw{\"o}ck}},\ }\href {\doibase 10.1016/j.aop.2010.09.012} {\bibfield
  {journal} {\bibinfo  {journal} {Ann. Phys.}\ }\textbf {\bibinfo {volume}
  {326}},\ \bibinfo {pages} {96} (\bibinfo {year} {2011})}\BibitemShut
  {NoStop}%
\bibitem [{\citenamefont {Luitz}\ \emph
  {et~al.}(2014{\natexlab{a}})\citenamefont {Luitz}, \citenamefont
  {Laflorencie},\ and\ \citenamefont {Alet}}]{luitz2014participation}%
  \BibitemOpen
  \bibfield  {author} {\bibinfo {author} {\bibfnamefont {D.~J.}\ \bibnamefont
  {Luitz}}, \bibinfo {author} {\bibfnamefont {N.}~\bibnamefont {Laflorencie}},
  \ and\ \bibinfo {author} {\bibfnamefont {F.}~\bibnamefont {Alet}},\ }\href
  {https://iopscience.iop.org/article/10.1088/1742-5468/2014/08/P08007}
  {\bibfield  {journal} {\bibinfo  {journal} {J. Stat. Mech.}\ }\textbf
  {\bibinfo {volume} {2014}},\ \bibinfo {pages} {P08007} (\bibinfo {year}
  {2014}{\natexlab{a}})}\BibitemShut {NoStop}%
\bibitem [{\citenamefont {St\'ephan}\ \emph {et~al.}(2009)\citenamefont
  {St\'ephan}, \citenamefont {Furukawa}, \citenamefont {Misguich},\ and\
  \citenamefont {Pasquier}}]{stephan2009shannon}%
  \BibitemOpen
  \bibfield  {author} {\bibinfo {author} {\bibfnamefont {J.-M.}\ \bibnamefont
  {St\'ephan}}, \bibinfo {author} {\bibfnamefont {S.}~\bibnamefont {Furukawa}},
  \bibinfo {author} {\bibfnamefont {G.}~\bibnamefont {Misguich}}, \ and\
  \bibinfo {author} {\bibfnamefont {V.}~\bibnamefont {Pasquier}},\ }\href
  {\doibase 10.1103/PhysRevB.80.184421} {\bibfield  {journal} {\bibinfo
  {journal} {Phys. Rev. B}\ }\textbf {\bibinfo {volume} {80}},\ \bibinfo
  {pages} {184421} (\bibinfo {year} {2009})}\BibitemShut {NoStop}%
\bibitem [{\citenamefont {St\'ephan}\ \emph {et~al.}(2010)\citenamefont
  {St\'ephan}, \citenamefont {Misguich},\ and\ \citenamefont
  {Pasquier}}]{stephan2009renyi}%
  \BibitemOpen
  \bibfield  {author} {\bibinfo {author} {\bibfnamefont {J.-M.}\ \bibnamefont
  {St\'ephan}}, \bibinfo {author} {\bibfnamefont {G.}~\bibnamefont {Misguich}},
  \ and\ \bibinfo {author} {\bibfnamefont {V.}~\bibnamefont {Pasquier}},\
  }\href {\doibase 10.1103/PhysRevB.82.125455} {\bibfield  {journal} {\bibinfo
  {journal} {Phys. Rev. B}\ }\textbf {\bibinfo {volume} {82}},\ \bibinfo
  {pages} {125455} (\bibinfo {year} {2010})}\BibitemShut {NoStop}%
\bibitem [{\citenamefont {Alcaraz}\ and\ \citenamefont
  {Rajabpour}(2013)}]{alcaraz2013}%
  \BibitemOpen
  \bibfield  {author} {\bibinfo {author} {\bibfnamefont {F.~C.}\ \bibnamefont
  {Alcaraz}}\ and\ \bibinfo {author} {\bibfnamefont {M.~A.}\ \bibnamefont
  {Rajabpour}},\ }\href {\doibase 10.1103/PhysRevLett.111.017201} {\bibfield
  {journal} {\bibinfo  {journal} {Phys. Rev. Lett.}\ }\textbf {\bibinfo
  {volume} {111}},\ \bibinfo {pages} {017201} (\bibinfo {year}
  {2013})}\BibitemShut {NoStop}%
\bibitem [{\citenamefont {St\'ephan}(2014)}]{stephan2014renyi}%
  \BibitemOpen
  \bibfield  {author} {\bibinfo {author} {\bibfnamefont {J.-M.}\ \bibnamefont
  {St\'ephan}},\ }\href {\doibase 10.1103/PhysRevB.90.045424} {\bibfield
  {journal} {\bibinfo  {journal} {Phys. Rev. B}\ }\textbf {\bibinfo {volume}
  {90}},\ \bibinfo {pages} {045424} (\bibinfo {year} {2014})}\BibitemShut
  {NoStop}%
\bibitem [{\citenamefont {Fradkin}\ and\ \citenamefont
  {Moore}(2006)}]{fradkin2006entanglement}%
  \BibitemOpen
  \bibfield  {author} {\bibinfo {author} {\bibfnamefont {E.}~\bibnamefont
  {Fradkin}}\ and\ \bibinfo {author} {\bibfnamefont {J.~E.}\ \bibnamefont
  {Moore}},\ }\href {\doibase 10.1103/PhysRevLett.97.050404} {\bibfield
  {journal} {\bibinfo  {journal} {Phys. Rev. Lett.}\ }\textbf {\bibinfo
  {volume} {97}},\ \bibinfo {pages} {050404} (\bibinfo {year}
  {2006})}\BibitemShut {NoStop}%
\bibitem [{\citenamefont {Hsu}\ and\ \citenamefont
  {Fradkin}(2010)}]{hsu2010universal}%
  \BibitemOpen
  \bibfield  {author} {\bibinfo {author} {\bibfnamefont {B.}~\bibnamefont
  {Hsu}}\ and\ \bibinfo {author} {\bibfnamefont {E.}~\bibnamefont {Fradkin}},\
  }\href {\doibase 10.1088/1742-5468/2010/09/P09004} {\bibfield  {journal}
  {\bibinfo  {journal} {J. Stat. Mech.}\ }\textbf {\bibinfo {volume} {2010}},\
  \bibinfo {pages} {P09004} (\bibinfo {year} {2010})}\BibitemShut {NoStop}%
\bibitem [{\citenamefont {Luitz}\ \emph
  {et~al.}(2014{\natexlab{b}})\citenamefont {Luitz}, \citenamefont {Alet},\
  and\ \citenamefont {Laflorencie}}]{luitz2014universal}%
  \BibitemOpen
  \bibfield  {author} {\bibinfo {author} {\bibfnamefont {D.~J.}\ \bibnamefont
  {Luitz}}, \bibinfo {author} {\bibfnamefont {F.}~\bibnamefont {Alet}}, \ and\
  \bibinfo {author} {\bibfnamefont {N.}~\bibnamefont {Laflorencie}},\ }\href
  {\doibase 10.1103/PhysRevLett.112.057203} {\bibfield  {journal} {\bibinfo
  {journal} {Phys. Rev. Lett.}\ }\textbf {\bibinfo {volume} {112}},\ \bibinfo
  {pages} {057203} (\bibinfo {year} {2014}{\natexlab{b}})}\BibitemShut
  {NoStop}%
\bibitem [{\citenamefont {Luitz}\ \emph
  {et~al.}(2014{\natexlab{c}})\citenamefont {Luitz}, \citenamefont {Plat},
  \citenamefont {Laflorencie},\ and\ \citenamefont
  {Alet}}]{luitz2014improving}%
  \BibitemOpen
  \bibfield  {author} {\bibinfo {author} {\bibfnamefont {D.~J.}\ \bibnamefont
  {Luitz}}, \bibinfo {author} {\bibfnamefont {X.}~\bibnamefont {Plat}},
  \bibinfo {author} {\bibfnamefont {N.}~\bibnamefont {Laflorencie}}, \ and\
  \bibinfo {author} {\bibfnamefont {F.}~\bibnamefont {Alet}},\ }\href {\doibase
  10.1103/PhysRevB.90.125105} {\bibfield  {journal} {\bibinfo  {journal} {Phys.
  Rev. B}\ }\textbf {\bibinfo {volume} {90}},\ \bibinfo {pages} {125105}
  (\bibinfo {year} {2014}{\natexlab{c}})}\BibitemShut {NoStop}%
\bibitem [{Note1()}]{Note1}%
  \BibitemOpen
  \bibinfo {note} {The same discussion holds for MPSs which are invariant under
  shift of $p$ sites, with $p>1$.}\BibitemShut {Stop}%
\bibitem [{Note2()}]{Note2}%
  \BibitemOpen
  \bibinfo {note} {This is a working hypothesis encoding ``typical behavior''
  of MPSs, and which simplifies our derivations. However, we do not expect it
  to be necessary, see also Appendix~\ref {sec:locality}.}\BibitemShut {Stop}%
\bibitem [{\citenamefont {Sachdev}(2011)}]{sachdev2011}%
  \BibitemOpen
  \bibfield  {author} {\bibinfo {author} {\bibfnamefont {S.}~\bibnamefont
  {Sachdev}},\ }\href@noop {} {\emph {\bibinfo {title} {Quantum phase
  transitions}}}\ (\bibinfo  {publisher} {Harvard University, Massachusetts},\
  \bibinfo {year} {2011})\BibitemShut {NoStop}%
\bibitem [{\citenamefont {Fishman}\ \emph {et~al.}(2022)\citenamefont
  {Fishman}, \citenamefont {White},\ and\ \citenamefont
  {Stoudenmire}}]{itensor}%
  \BibitemOpen
  \bibfield  {author} {\bibinfo {author} {\bibfnamefont {M.}~\bibnamefont
  {Fishman}}, \bibinfo {author} {\bibfnamefont {S.}~\bibnamefont {White}}, \
  and\ \bibinfo {author} {\bibfnamefont {E.}~\bibnamefont {Stoudenmire}},\
  }\href@noop {} {\bibfield  {journal} {\bibinfo  {journal} {SciPost Physics
  Codebases}\ ,\ \bibinfo {pages} {004}} (\bibinfo {year} {2022})}\BibitemShut
  {NoStop}%
\bibitem [{Note3()}]{Note3}%
  \BibitemOpen
  \bibinfo {note} {For $|h|<1$ the GS is two-fold degenerate for $N\to \infty
  $. Following Ref.~\cite {oliviero2022magic}, we focused on the exact GS at
  finite $N$, which is symmetric with respect to the $\protect \mathbb {Z}_2$
  symmetry $Z=\DOTSB \prod@ \slimits@ _j\sigma ^z_j$, but we expect that the
  density of magic is the same for the two short-range correlated
  symmetry-broken GSs.}\BibitemShut {Stop}%
\bibitem [{\citenamefont {Osterloh}\ \emph {et~al.}(2002)\citenamefont
  {Osterloh}, \citenamefont {Amico}, \citenamefont {Falci},\ and\ \citenamefont
  {Fazio}}]{osterloh2002scaling}%
  \BibitemOpen
  \bibfield  {author} {\bibinfo {author} {\bibfnamefont {A.}~\bibnamefont
  {Osterloh}}, \bibinfo {author} {\bibfnamefont {L.}~\bibnamefont {Amico}},
  \bibinfo {author} {\bibfnamefont {G.}~\bibnamefont {Falci}}, \ and\ \bibinfo
  {author} {\bibfnamefont {R.}~\bibnamefont {Fazio}},\ }\href {\doibase
  10.1038/416608a} {\bibfield  {journal} {\bibinfo  {journal} {Nature}\
  }\textbf {\bibinfo {volume} {416}},\ \bibinfo {pages} {608} (\bibinfo {year}
  {2002})}\BibitemShut {NoStop}%
\bibitem [{\citenamefont {Calabrese}\ and\ \citenamefont
  {Cardy}(2004)}]{calabrese2004entanglement}%
  \BibitemOpen
  \bibfield  {author} {\bibinfo {author} {\bibfnamefont {P.}~\bibnamefont
  {Calabrese}}\ and\ \bibinfo {author} {\bibfnamefont {J.}~\bibnamefont
  {Cardy}},\ }\href {\doibase 10.1088/1742-5468/2004/06/P06002} {\bibfield
  {journal} {\bibinfo  {journal} {J. Stat. Mech.}\ }\textbf {\bibinfo {volume}
  {2004}},\ \bibinfo {pages} {P06002} (\bibinfo {year} {2004})}\BibitemShut
  {NoStop}%
\bibitem [{\citenamefont {Amico}\ \emph {et~al.}(2008)\citenamefont {Amico},
  \citenamefont {Fazio}, \citenamefont {Osterloh},\ and\ \citenamefont
  {Vedral}}]{amico2008entanglement}%
  \BibitemOpen
  \bibfield  {author} {\bibinfo {author} {\bibfnamefont {L.}~\bibnamefont
  {Amico}}, \bibinfo {author} {\bibfnamefont {R.}~\bibnamefont {Fazio}},
  \bibinfo {author} {\bibfnamefont {A.}~\bibnamefont {Osterloh}}, \ and\
  \bibinfo {author} {\bibfnamefont {V.}~\bibnamefont {Vedral}},\ }\href
  {\doibase 10.1103/RevModPhys.80.517} {\bibfield  {journal} {\bibinfo
  {journal} {Rev. Mod. Phys.}\ }\textbf {\bibinfo {volume} {80}},\ \bibinfo
  {pages} {517} (\bibinfo {year} {2008})}\BibitemShut {NoStop}%
\bibitem [{\citenamefont {Calabrese}\ and\ \citenamefont
  {Cardy}(2009)}]{calabrese2009entanglement}%
  \BibitemOpen
  \bibfield  {author} {\bibinfo {author} {\bibfnamefont {P.}~\bibnamefont
  {Calabrese}}\ and\ \bibinfo {author} {\bibfnamefont {J.}~\bibnamefont
  {Cardy}},\ }\href {\doibase 10.1088/1751-8113/42/50/504005} {\bibfield
  {journal} {\bibinfo  {journal} {J. Phys. A: Math. Theor.}\ }\textbf {\bibinfo
  {volume} {42}},\ \bibinfo {pages} {504005} (\bibinfo {year}
  {2009})}\BibitemShut {NoStop}%
\bibitem [{\citenamefont {Sierant}\ and\ \citenamefont
  {Turkeshi}(2022)}]{sierant2022universal}%
  \BibitemOpen
  \bibfield  {author} {\bibinfo {author} {\bibfnamefont {P.}~\bibnamefont
  {Sierant}}\ and\ \bibinfo {author} {\bibfnamefont {X.}~\bibnamefont
  {Turkeshi}},\ }\href {\doibase 10.1103/PhysRevLett.128.130605} {\bibfield
  {journal} {\bibinfo  {journal} {Phys. Rev. Lett.}\ }\textbf {\bibinfo
  {volume} {128}},\ \bibinfo {pages} {130605} (\bibinfo {year}
  {2022})}\BibitemShut {NoStop}%
\bibitem [{\citenamefont {Nelder}\ and\ \citenamefont
  {Mead}(1965)}]{nelder1965simplex}%
  \BibitemOpen
  \bibfield  {author} {\bibinfo {author} {\bibfnamefont {J.~A.}\ \bibnamefont
  {Nelder}}\ and\ \bibinfo {author} {\bibfnamefont {R.}~\bibnamefont {Mead}},\
  }\href@noop {} {\bibfield  {journal} {\bibinfo  {journal} {The computer
  journal}\ }\textbf {\bibinfo {volume} {7}},\ \bibinfo {pages} {308} (\bibinfo
  {year} {1965})}\BibitemShut {NoStop}%
\bibitem [{\citenamefont {Ellison}\ \emph {et~al.}(2021)\citenamefont
  {Ellison}, \citenamefont {Kato}, \citenamefont {Liu},\ and\ \citenamefont
  {Hsieh}}]{ellison2021symmetry}%
  \BibitemOpen
  \bibfield  {author} {\bibinfo {author} {\bibfnamefont {T.~D.}\ \bibnamefont
  {Ellison}}, \bibinfo {author} {\bibfnamefont {K.}~\bibnamefont {Kato}},
  \bibinfo {author} {\bibfnamefont {Z.-W.}\ \bibnamefont {Liu}}, \ and\
  \bibinfo {author} {\bibfnamefont {T.~H.}\ \bibnamefont {Hsieh}},\ }\href
  {\doibase 10.22331/q-2021-12-28-612} {\bibfield  {journal} {\bibinfo
  {journal} {Quantum}\ }\textbf {\bibinfo {volume} {5}},\ \bibinfo {pages}
  {612} (\bibinfo {year} {2021})}\BibitemShut {NoStop}%
\bibitem [{\citenamefont {Silvi}\ \emph {et~al.}(2019)\citenamefont {Silvi},
  \citenamefont {Tschirsich}, \citenamefont {Gerster}, \citenamefont
  {Jünemann}, \citenamefont {Jaschke}, \citenamefont {Rizzi},\ and\
  \citenamefont {Montangero}}]{silvi2019tensor}%
  \BibitemOpen
  \bibfield  {author} {\bibinfo {author} {\bibfnamefont {P.}~\bibnamefont
  {Silvi}}, \bibinfo {author} {\bibfnamefont {F.}~\bibnamefont {Tschirsich}},
  \bibinfo {author} {\bibfnamefont {M.}~\bibnamefont {Gerster}}, \bibinfo
  {author} {\bibfnamefont {J.}~\bibnamefont {Jünemann}}, \bibinfo {author}
  {\bibfnamefont {D.}~\bibnamefont {Jaschke}}, \bibinfo {author} {\bibfnamefont
  {M.}~\bibnamefont {Rizzi}}, \ and\ \bibinfo {author} {\bibfnamefont
  {S.}~\bibnamefont {Montangero}},\ }\href {\doibase
  10.21468/SciPostPhysLectNotes.8} {\bibfield  {journal} {\bibinfo  {journal}
  {SciPost Phys. Lect. Notes}\ ,\ \bibinfo {pages} {8}} (\bibinfo {year}
  {2019})}\BibitemShut {NoStop}%
\bibitem [{\citenamefont {Evans}\ and\ \citenamefont
  {H{\o}egh-Krohn}(1977)}]{evans1977spectral}%
  \BibitemOpen
  \bibfield  {author} {\bibinfo {author} {\bibfnamefont {D.~E.}\ \bibnamefont
  {Evans}}\ and\ \bibinfo {author} {\bibfnamefont {R.}~\bibnamefont
  {H{\o}egh-Krohn}},\ }\href {\doibase 10.1112/jlms/s2-17.2.345} {\bibfield
  {journal} {\bibinfo  {journal} {J. London Math. Soc.}\ }\textbf {\bibinfo
  {volume} {17}},\ \bibinfo {pages} {345} (\bibinfo {year} {1977})}\BibitemShut
  {NoStop}%
\end{thebibliography}%

\end{document}